\documentclass[a4paper,11pt]{article}
\pdfoutput=1 % to allow compilation in JCAP
\usepackage{jcappub}
\usepackage{amsmath}
\usepackage{bm}
\usepackage{graphicx}
\usepackage{enumitem}
\usepackage{geometry}
\usepackage{xspace}
\usepackage{pdflscape}
\usepackage{dsfont}
\usepackage{tensor}
\usepackage{comment}
\usepackage{soul}

\usepackage{mathtools}

\usepackage{relsize}

\allowdisplaybreaks % allows long equations to be displayed over several pages

\makeatletter
\gdef\@fpheader{}
\g@addto@macro\bfseries{\boldmath}
\makeatother

\newcommand{\zetac}{\zeta_{\mathrm{c}}}

\newcommand{\dd}{\mathrm{d}}
\newcommand{\calN}{\mathcal{N}}

\newcommand{\deep}{deep\xspace}
\newcommand{\shallow}{shallow\xspace}

\newcommand{\Eq}[1]{Eq.~(\ref{#1})}
\newcommand{\Eqs}[1]{Eqs.~(\ref{#1})}
\newcommand{\Fig}[1]{Fig.~{\ref{#1}}}
\newcommand{\Figs}[1]{Figs.~{\ref{#1}}}
\newcommand{\Refa}[1]{Ref.~{\cite{#1}}}
\newcommand{\Refs}[1]{Refs.~{\cite{#1}}}
\newcommand{\Sec}[1]{Sec.~\ref{#1}}

\newcommand{\App}[1]{Appendix~\ref{#1}}

\newcommand{\ie}{\textsl{i.e.}~}

\newcommand{\eg}{\textsl{e.g.}\xspace}

\newcommand{\phiuv}{\phi_{\text{uv}}}

\newcommand{\N}{\mathcal{N}}

\let\oldsqrt\sqrt
\def\sqrt{\mathpalette\DHLhksqrt}
\def\DHLhksqrt#1#2{%
\setbox0=\hbox{$#1\oldsqrt{#2\,}$}\dimen0=\ht0
\advance\dimen0-0.2\ht0
\setbox2=\hbox{\vrule height\ht0 depth -\dimen0}%
{\box0\lower0.4pt\box2}}

\newcommand{\hyp}{\,_1\mathrm{F}_1}
\newcommand{\mean}[1]{\left\langle #1 \right\rangle}
\newcommand{\order}[1]{\mathcal{O}\!\left(#1\right)}

\DeclareMathOperator{\erf}{erf}

\DeclareMathOperator{\erfi}{erfi}
\DeclareMathOperator{\Lag}{L}

\DeclareMathOperator{\Her}{H}

\newcommand{\ee}{e}

\newcommand{\sss}[1]{{\scriptscriptstyle{#1}}}

\newcommand{\uPl}{\mathrm{Pl}}

\newcommand{\uend}{\mathrm{end}}

\newcommand{\uc}{\mathrm{c}}

\newcommand{\usssPl}{\sss{\uPl}}

\newcommand{\uNL}{\mathrm{NL}}

\newcommand{\Mp}{M_\usssPl}

\newcommand{\fnl}{f_\uNL}

\newcommand{\efolds}{$e$-folds\xspace}

\newcommand{\beq}{\begin{equation}}
\newcommand{\eeq}{\end{equation}}
\newcommand{\bea}{\begin{equation}\begin{aligned}}
\newcommand{\eea}{\end{aligned}\end{equation}}

\newlength{\wsingfig}
\newlength{\wdblefig}
\newlength{\wquadfig}
\newlength{\wtriplefig}
\newcommand{\deflen}[2]{%      
    \expandafter\newlength\csname #1\endcsname
    \expandafter\setlength\csname #1\endcsname{#2}%
}

\newcommand{\derhyp}{\prime}

\subheader{}

\title{ Primordial black holes from stochastic tunnelling}

\author[a,b]{Chiara Animali,}
\author[b]{Vincent Vennin}

\affiliation[a]{Dipartimento di Fisica, Universit\`a di Pisa, Largo B. Pontecorvo 3, 56127 Pisa, Italy, and Istituto Nazionale di Fisica Nucleare, Sezione di Pisa, Pisa, Italy}

\affiliation[b]{Laboratoire de Physique de l'\'Ecole Normale Sup\'erieure, ENS, CNRS, Universit\'e PSL, Sorbonne Universit\'e, Universit\'e Paris Cit\'e, F-75005 Paris, France}

\emailAdd{chiara.animali@ens.fr}
\emailAdd{vincent.vennin@ens.fr}

\date{today}

\begin{document}
\sloppy

%------------
%ABSTRACT
%------------
\abstract{
If the inflaton gets trapped in a local minimum of its potential shortly before the end of inflation, it escapes by building up quantum fluctuations in a process known as stochastic tunnelling. In this work we study cosmological fluctuations produced in such a scenario, and how likely they are to form Primordial Black Holes (PBHs). This is done by using the stochastic-$\delta N$ formalism, which allows us to reconstruct the highly non-Gaussian tails of the distribution function of the number of \efolds spent in the false-vacuum state. We explore two different toy models, both analytically and numerically, in order to identify which properties do or do not depend on the details of the false-vacuum profile. We find that when the potential barrier is small enough compared to its width, $\Delta V/V < \Delta\phi^2/\Mp^2$, the potential can be approximated as being flat between its two local extrema, so results previously obtained in a ``flat quantum well'' apply. 
Otherwise, when $\Delta  V/V < V/\Mp^4$, the PBH abundance depends exponentially on the height of the potential barrier, and when $\Delta  V/V > V/\Mp^4$ it depends super-exponentially (\ie as the exponential of an exponential) on the barrier height. In that later case PBHs are massively produced.
This allows us to quantify how much flat inflection points need to be fine-tuned. 
In a deep false vacuum, we also find that slow-roll violations are typically encountered unless the potential is close to linear. This motivates further investigations to generalise our approach to non--slow-roll setups. 
}

%\keywords{physics of the early universe, inflation, primordial black holes}

%\arxivnumber{XXXX.XXXX}

\maketitle

%------------
%SECTION: INTRODUCTION
%------------

\section{Introduction and motivations}
\label{sec:Introduction}

Cosmic inflation is, to date, the simplest and phenomenologically most successful paradigm to describe the early universe~\cite{Starobinsky:1980te, Sato:1980yn, Guth:1980zm, Linde:1981mu, Albrecht:1982wi, Linde:1983gd}. Other than solving a number of puzzles in the standard Big Bang cosmology, this early phase of almost exponential expansion also provides the seeds for all cosmological structures in our universe, through the parametric amplification of quantum vacuum fluctuations~\cite{Starobinsky:1979ty,Mukhanov:1981xt, Starobinsky:1982ee,Guth:1982ec,Bardeen:1983qw}. These cosmological perturbations can be observed in the Cosmic Microwave Background (CMB) anisotropies~\cite{Planck:2018vyg, Planck:2018nkj, Planck:2019kim}, and in the Large-Scale Structures (LSS) of the universe~\cite{Amendola:2016saw}.
They are predicted to be almost scale-invariant, quasi-Gaussian and quasi-adiabatic, which is in excellent agreement with current observations~\cite{Planck:2018nkj}.

However, we are still far from having a complete picture of the early universe. CMB and LSS observations only give access to a restricted range of physical wavelengths, which emerged from the Hubble radius during a short period of about 7 \efolds, over the $\sim 50$ \efolds required to account for the observable universe. At these scales, perturbations are constrained to be small, at the level of $\zeta \simeq 10^{-5}$, where $\zeta$ is the so-called curvature perturbation~\cite{Malik:2008im}. Even though this is enough to indicate that the inflationary potential could be of the plateau type around $50$ \efolds before the end of inflation (at least in the minimal setting of single-field slow-roll inflation)~\cite{Martin:2013tda, Martin:2013nzq}, the lack of observational constraints at small scales makes the reconstruction of the inflationary potential close to the end of inflation still elusive. On the one hand, this calls for new observational windows at small scales; on the other hand, this prompts us to keep an open view about possible deviations from ``vanilla inflation'' outside the constrained range~\cite{Karam:2022nym}, and to investigate possible phenomenological consequences that might be looked for in those new windows. 

\subsection*{False vacuum}
\begin{figure}[t]
\centering 
\includegraphics[width=.49\textwidth]{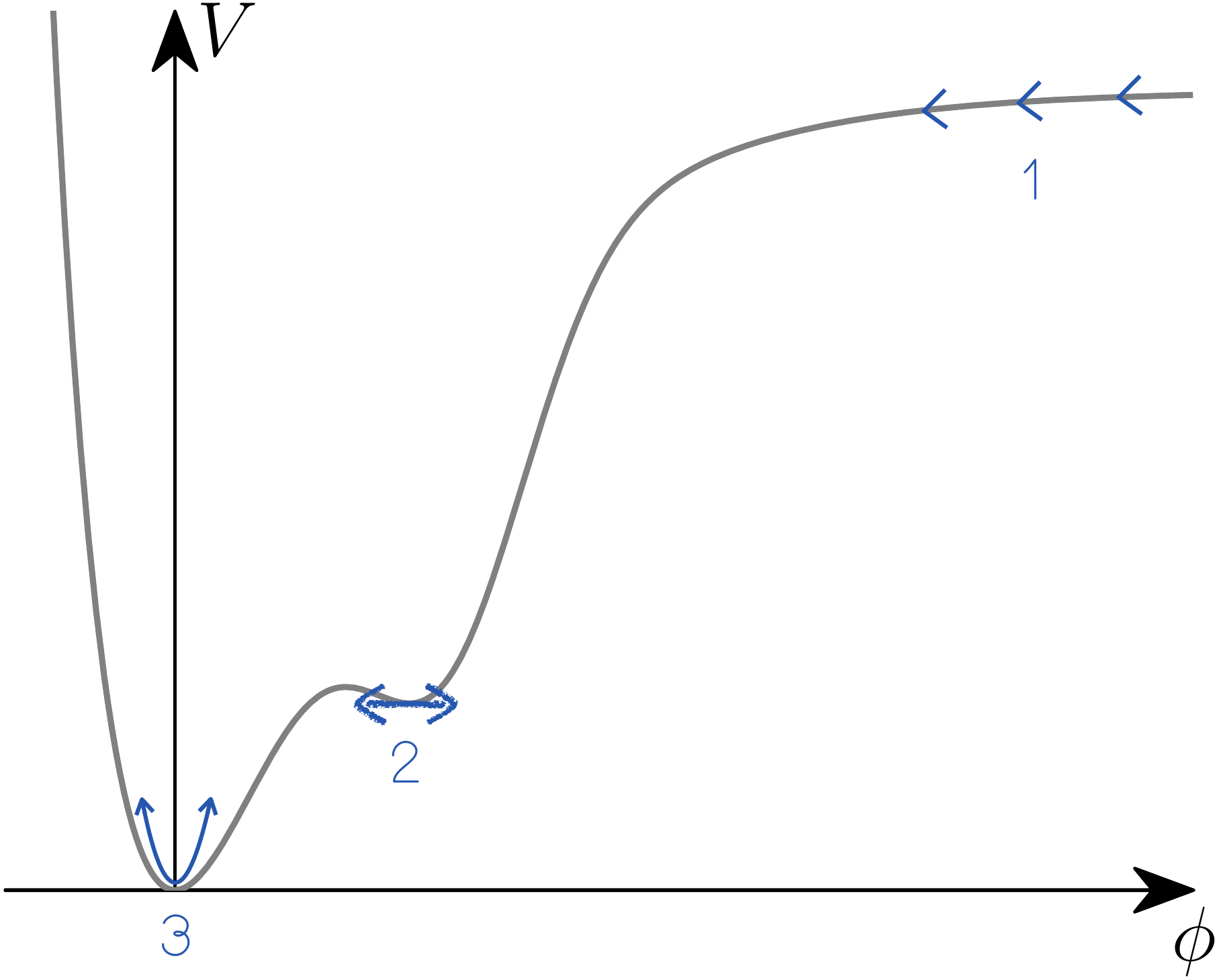}
 \caption{Sketch of the inflationary potential considered in this work. The large scales observed in the CMB and the LSS exit the Hubble radius at large-field value, where the potential is of the plateau type (1). The inflaton $\phi$ then falls in a false vacuum state, \ie a local minimum, from which it escapes through quantum fluctuations (2). It finally reaches the true vacuum state, around which it oscillates during the preheating phase (3).}
 \label{fig:sketch}
\end{figure}
Among the panorama of possible features that could alter the inflationary potential at small scales, in this work we consider the possibility of a false vacuum state, \ie a local minimum that is up-shifted compared to the global vacuum in which (p)reheating takes place. The situation is depicted in \Fig{fig:sketch}. 
This is somehow reminiscent of the ``old inflation'' proposal~\cite{Guth:1980zm}, where inflation occurs while the universe is trapped in a false vacuum state, from which a graceful exit to the true vacuum is enabled by quantum tunnelling~\cite{Coleman:1977py, Coleman:1980aw}. It was  then realised that reheating through percolation of true-vacuum bubbles was challenging in this setup~\cite{Guth:1982pn}. This led to the ``new inflation'' proposal~\cite{Linde:1981mu, Albrecht:1982wi}, in which inflation is driven by a scalar field $\phi$ along a smooth potential $V(\phi)$, and terminates by violation of the slow-roll conditions. The situation depicted in \Fig{fig:sketch} may thus be seen as a hybrid setup mixing old and new inflation (see \Refs{Dvali:2003vv, Kitajima:2019ibn} for other setups combining the two mechanisms). 

False vacua may appear naturally in various high-energy constructions of inflationary potentials: when embedded in supersymmetry or supergravity constructions, plateau-like potentials at CMB scales often include features like local minima at smaller scales~\cite{Geller:2022nkr}. Moreover, in models yielding inflection points in the potentials, as usually encountered in string- or supersymmetry-inspired constructions~\cite{Gherghetta:1995dv, Enqvist:2003gh, Allahverdi:2006iq, Lyth:2006ec} (and as often studied in the context of primordial black hole production~\cite{Kawasaki:2016pql,Ezquiaga:2017fvi, Garcia-Bellido:2017mdw}), the breaking of the flat-inflection point condition through radiative corrections~\cite{Allahverdi:2006wt, Allahverdi:2008bt, Chatterjee:2011qr} may create an additional local minimum, depending on its sign. Finally, false vacua are found in more specific scenarios such as the critical Higgs inflationary model~\cite{Hamada:2014wna,Bezrukov:2014bra}.

When the inflaton encounters a false vacuum, there are two ways it can climb up the potential barrier and reach the global minimum of the potential, close to which inflation ends. This first possibility, as recently investigated in \Refs{Inomata:2021tpx, Cai:2022erk, Geller:2022nkr, Gu:2022pbo}, is that the field's classical velocity is large enough to overshoot the local minimum. In that case, the slow-roll conditions are necessarily violated, which leads to a sharp increase in the amplitude of curvature perturbations that may later collapse into primordial black holes (PBHs)~\cite{Hawking:1971ei, Carr:1974nx, Carr:1975qj}. The second possibility is that the inflaton gets trapped in the false vacuum. In that case, quantum fluctuations jiggle the inflaton and after some time they shall push it outwards. In that case, large fluctuations have to build up, and as a consequence one may also expect PBHs to form. 

The goal of this paper is to put this statement under closer scrutiny, and investigate how quantum diffusion proceeds in a false-vacuum state during inflation. The reason is that, while non-linear structure formation makes primordial fluctuations of small amplitude difficult to reconstruct at small scales, the presence of PBHs is a distinct feature that can be specifically looked for (and, if not detected, at least constrained). Therefore, they constitute an important window into small-scales fluctuations of large-enough amplitude, hence into the potential presence of a false-vacuum state during inflation.

\subsection*{Stochastic-$\delta N$ formalism}

During inflation, quantum diffusion can be described by means of the stochastic-inflation formalism~\cite{Starobinsky:1982ee,Starobinsky:1986fx}, where small-scale fluctuations behave as a random noise acting on the large-scale evolution as they cross out the Hubble radius. These ``quantum kicks'' make the inflaton wriggle away from the false vacuum, a process known as stochastic tunnelling (note that it is different from standard quantum tunnelling that proceeds below the potential barrier and for which there is no classical description). Stochastic tunnelling has been studied in various contexts in \Refs{Ellis:1990bv, Linde:1991sk, Espinosa:2007qp, Tolley:2008qv, Hook:2014uia, Kearney:2015vba, Espinosa:2015qea, East:2016anr, Noorbala:2018zlv, Bramberger:2019zks, Kitajima:2019ibn, Fumagalli:2019ohr, Hertzberg:2020tqa, Camargo-Molina:2022ord}, and here we investigate how this mechanism affects cosmological perturbations.
This can be done by following the stochastic-$\delta N$ approach, which we now briefly summarise in the case of a single field $\phi$ slowly rolling on a potential $V(\phi)$ (see \eg \Refa{Vennin:2020kng} for a more extensive review).

On the slow-roll attractor, the long-wavelength part of the inflaton is driven by the Langevin equation
\beq
\label{eq:Langevin}
\frac{\dd \phi}{\dd N}=-\frac{V^\prime(\phi)}{3 H^2(\phi)}+\frac{H(\phi)}{2\pi}\, \xi\,,
\eeq
where $N=\ln(a)$ is the number of \efolds~\cite{Finelli:2008zg, Pattison:2019hef} with $a$ the scale factor, a prime denotes a derivative with respect to $\phi$, $H=\dot{a}/a$ is the Hubble parameter where a dot denotes derivation with respect to cosmic time, and $\xi$ is a white Gaussian noise with vanishing mean and unit variance, \ie $\langle \xi(N) \xi(N')\rangle=\delta(N-N')$. This noise describes the inflow of small wavelength scales as the universe expansion stretches them into the long-wavelength sector, and hereafter $\langle\cdot\rangle$ denotes stochastic average. At leading order in slow roll, $H^2\simeq V/(3\Mp^2)$ where $\Mp$ is the reduced Planck mass, because of Friedmann equation. This Langevin equation can then be turned into a Fokker-Planck equation for the probability density function (PDF), $P(\phi,N)$, associated with the field value at time $N$,
\bea
\frac{\partial }{\partial N} P(\phi,N) = \Mp^2 \frac{\partial}{\partial \phi}\left[\frac{v'(\phi)}{v(\phi)}P(\phi,N)\right]+\Mp^2\frac{\partial^2}{\partial \phi^2}\left[v(\phi) P(\phi,N)\right]\,,
\eea
where for convenience we have introduced the rescaled potential $v=V/(24\pi^2\Mp^4)$. Starting from a certain initial field value $\phi$, let $\N$ denote the number of \efolds that is realised until inflation ends. This is a random quantity, since it is different for each realisation of the stochastic process~\eqref{eq:Langevin}. It is therefore endowed with a distribution function $P(\N,\phi)$, which can be shown to follow the adjoint Fokker-Planck equation~\cite{Vennin:2015hra,Pattison:2017mbe}, namely 
\bea
\label{eq:adjoint:FP}
\frac{\partial }{\partial \N} P(\N,\phi) = -\Mp^2\frac{v'(\phi)}{v(\phi)} \frac{\partial}{\partial \phi} P(\N,\phi) +\Mp^2v(\phi)\frac{\partial^2}{\partial \phi^2} P(\N,\phi).
\eea
This equation needs to be solved with the boundary condition $P(\N,\phi_\uend)=\delta(\N)$, assuming that inflation ends at $\phi_\uend$ (an additional boundary condition is sometimes needed at large-field values~\cite{Assadullahi:2016gkk,Vennin:2016wnk}). The statistics of cosmological perturbations can then be extracted using the $\delta N$ formalism~\cite{Sasaki:1995aw, Sasaki:1998ug, Lyth:2004gb, Lyth:2005fi}, which states that on super-Hubble scales, the curvature perturbation $\zeta$ is related to the integrated local amount of expansion of a homogeneous patch,
treated as a separate universe~\cite{Salopek:1990jq, Sasaki:1995aw, Wands:2000dp, Lyth:2003im, Rigopoulos:2003ak, Lyth:2005fi, Pattison:2019hef, Artigas:2021zdk}, \ie 
\bea
\label{eq:deltaN:zeta}
\zeta(\bm{x})=\calN(\bm{x})-\mean{\calN}\, .
\eea
In this expression, which is valid even at the non-perturbative level, $\zeta$ is coarsed-grained at the Hubble radius at the end of inflation, but the statistics of the curvature perturbation (and of related quantities such as the density contrast or the compaction function) when coarse-grained at arbitrary scales can be inferred using the techniques introduced in \Refa{Tada:2021zzj}. In this way, solving the first-passage time problem described by \Eq{eq:adjoint:FP} allows one to reconstruct the statistics of cosmological fluctuations on large scales, by taking into account the backreaction of small-scales quantum diffusion in a non-perturbative way. This is the so-called stochastic-$\delta N$ program~\cite{Enqvist:2008kt, Fujita:2013cna, Vennin:2015hra}, which we intend to apply to the false-vacuum setup in this work.

Let us stress that the above equations are written in the slow-roll regime, which at the classical level assumes that the acceleration term $\ddot{\phi}$ is subdominant in the Klein-Gordon equation $\ddot{\phi}+3H\dot{\phi}+V'=0$. One may be concerned that this condition cannot be satisfied around a local minimum of the potential, since there $V'=0$. However, slow roll being a dynamical attractor, if the potential function satisfies the slow-roll conditions then the system does not leave the attractor even when approaching a local minimum.\footnote{For explicitness, let us expand $V\simeq V_0+m^2\phi^2/2$ around a local minimum located at $\phi=0$. Upon linearising the Klein-Gordon and Friedmann equations around the phase-space point $(\phi=0,\dot{\phi}=0)$, in the regime $m^2\ll H_0^2= V_0/(3\Mp^2)$, one finds that $\phi(t)$ is attracted towards the solution $\phi\propto \ee^{-m^2 t/(3 H_0)}$, which is such that $3H\dot{\phi}\simeq -V'(\phi)$ and $\ddot{\phi}\simeq m^2/(9H_0^2) V'(\phi) \ll V'(\phi)$, and which therefore corresponds to the slow-roll attractor.} In other words, although it is true that $V'$ decreases to $0$ when reaching the local minimum, so do $3H\dot{\phi}$ and $\ddot{\phi}$, at a rate such that the acceleration term remains negligible. Our use of the slow-roll approximation is therefore fully justified, as long as one makes sure that the potential function satisfies the slow-roll conditions all along, which we will carefully check in what follows.

The rest of the paper is organised as follows. In \Sec{sec:Minimum} we solve the first-passage time problem in two toy-model potentials that feature a false vacuum: a linear potential with negative slope and a more refined quadratic-piecewise potential. Our goal is to determine which features are generic of false-vacuum models and which depend on its detailed properties. In \Sec{sec:PBHs}, we then derive estimates for the abundance of PBHs in these models. This allows us to constrain the shape of the local minimum, \ie its width, height and depth, from the existing upper bounds on PBHs. Finally, we provide concluding remarks in \Sec{sec:Conclusion}.

%------------
%SUBSECTION: Single field slow-roll inflation with a local minimum
%------------

\section{Statistics of fluctuations in a false vacuum}
\label{sec:Minimum}

In this section, we apply the stochastic-$\delta N$ program reviewed in \Sec{sec:Introduction} to the false-vacuum region of models such as the potential depicted in \Fig{fig:sketch}. This will lead us to investigate the production of PBHs in such models in \Sec{sec:PBHs}. Since PBHs arise from large density fluctuations, their abundance is driven by the tail of the distribution functions of cosmological perturbations. We thus start by explaining how such tails can be reconstructed, following the approach developed in \Refs{Pattison:2017mbe, Ezquiaga:2019ftu}.

\subsection{Tail reconstruction}

The distribution function of the curvature perturbation can be obtained by solving the adjoint Fokker-Planck equation~\eqref{eq:adjoint:FP}. This equation being a linear partial differential equation, it is convenient to introduce the characteristic function 
\beq\label{chi}
\chi(\phi,t)\equiv \mean {e^{i t \calN(\phi)}}=\int_{-\infty}^{\infty} e^{i t \calN} P (\calN,\phi) \dd \calN\,,
\eeq
which depends on a dummy variable $t$ and on the initial field configuration $\phi$. It is clear that the characteristic function is nothing but the inverse Fourier transform of the PDF, which can thus be obtained by  Fourier transforming
\beq\label{invFT}
P(\calN,\phi)=\frac{1}{2 \pi} \int_{-\infty}^{\infty} e^{-i t \N}\chi (t,\phi) \dd t\,.
\eeq
Moreover, by inserting \Eq{invFT} into the adjoint Fokker-Planck equation~\eqref{eq:adjoint:FP}, one obtains an ordinary differential equation for the characteristic function
\beq
\label{eq:adjoint:FP:chi}
-i t \chi(t,\phi)=-\Mp^2 \frac{v'(\phi)}{v(\phi)} \frac{\partial}{\partial \phi}\chi(t,\phi)+\Mp^2 v(\phi) \frac{\partial^2}{\partial \phi^2} \chi(t,\phi)\, ,
\eeq
with the boundary condition $\chi(t,\phi_\uend)=1$.
The procedure is therefore the following: solve \Eq{eq:adjoint:FP:chi} to compute the characteristic function, and infer the PDF using \Eq{invFT}.

This second step can be performed explicitly if the characteristic function can be decomposed in a pole expansion
\beq
\chi(t,\phi)=\sum_n \frac{a_n(\phi)}{\Lambda_n-i t}+g(t,\phi)\,,
\eeq
where $\Lambda_n$ are positive numbers that do not depend on $\phi$, $a_n$ are real functions of $\phi$ and $g(t,\phi)$ is a regular function of $t$~\cite{Ezquiaga:2019ftu}. Because of the residue theorem, the PDF thus reads
\beq\label{general_PDF}
P(\N,\phi)=\sum_n a_n(\phi) e^{-\Lambda_n\,\N}\, ,
\eeq
where we assume that the $\Lambda_n$ have been ordered, $0<\Lambda_0 <\Lambda_1 < \cdots \Lambda_n$. This may be viewed as a tail expansion, since terms of higher $n$ are more strongly suppressed at larger $\N$. Therefore, on the tails, the PDF is dominated by the first poles only. The main task becomes to find the zeros of the inverse characteristic function, $\Lambda_n$, and extract the residues by evaluating 
\bea
\label{residues}
a_n(\phi)=-i \left[\frac{\partial}{\partial t}\chi ^{-1}(t=-i\Lambda_n,\phi)\right]^{-1}\,.
\eea
When the location of the poles $\Lambda_n$ are known only approximately, a more convenient expression can be obtained by decomposing
\bea
\chi(t,\phi)=\frac{\chi_{\mathrm{num}}(t,\phi)}{\chi_{\mathrm{den}}(t)}\, ,
\eea
where the denominator $\chi_{\mathrm{den}}$ does not depend on $\phi$ since, as mentioned above, the poles are independent of the initial condition $\phi$. In this form, \Eq{residues} leads to 
\bea
\label{residues:alternative}
a_n(\phi)=- i\frac{\chi_{\mathrm{num}}(-i\Lambda_n,\phi)}{\frac{\partial}{\partial t}\chi_{\mathrm{den}}(-i\Lambda_n)}\, ,
\eea
where we have used that $\chi_{\mathrm{den}}(-i\Lambda_n)=0$ (if $\Lambda_n$ is known only approximately, this condition cannot be enforced exactly in \Eq{residues}, which then makes \Eq{residues:alternative} more accurate). 

In passing, let us stress that \Eq{general_PDF} indicates that the tail of the PDF for $\zeta$ has an exponential, rather than Gaussian, fall-off behaviour. This has strong consequences for the formation of extreme objects such as PBHs, since exponential tails are much heavier, but also for galaxy and structure formation in general~\cite{Ezquiaga:2022qpw}. This type of non-Gaussianities cannot be captured by perturbative parametrisations (such as the $\fnl$ expansion~\cite{Gangui:1993tt}), which only account for small deviations from Gaussian statistics around the maximum of the PDF~\cite{Byrnes:2012yx, Passaglia:2018ixg, Atal:2018neu} and not on its tail.

\subsection{Linear toy model}
\label{subsec:Linear}
\begin{figure}[t]
\centering 
\includegraphics[width=.49\textwidth]{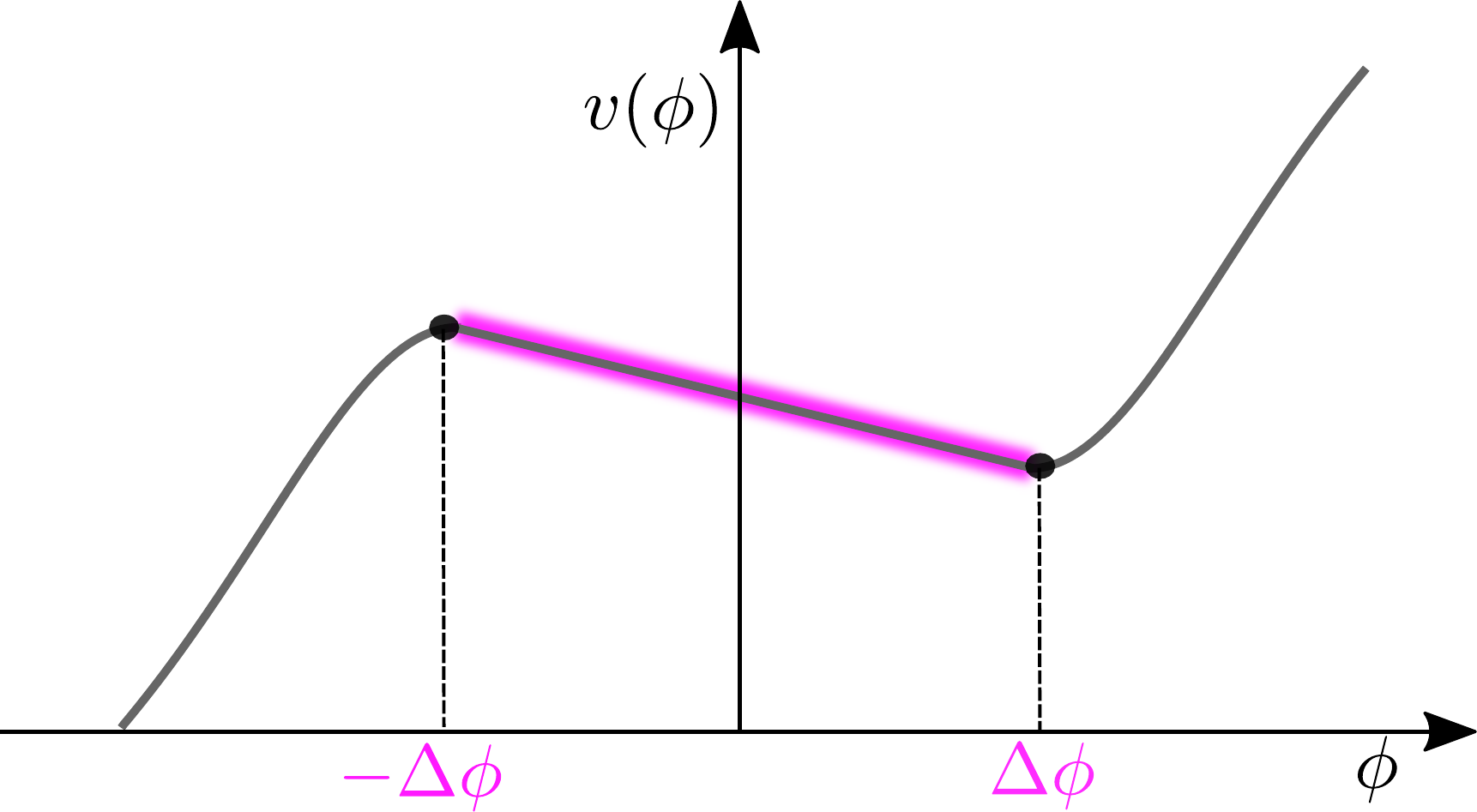}
\includegraphics[width=.49\textwidth]{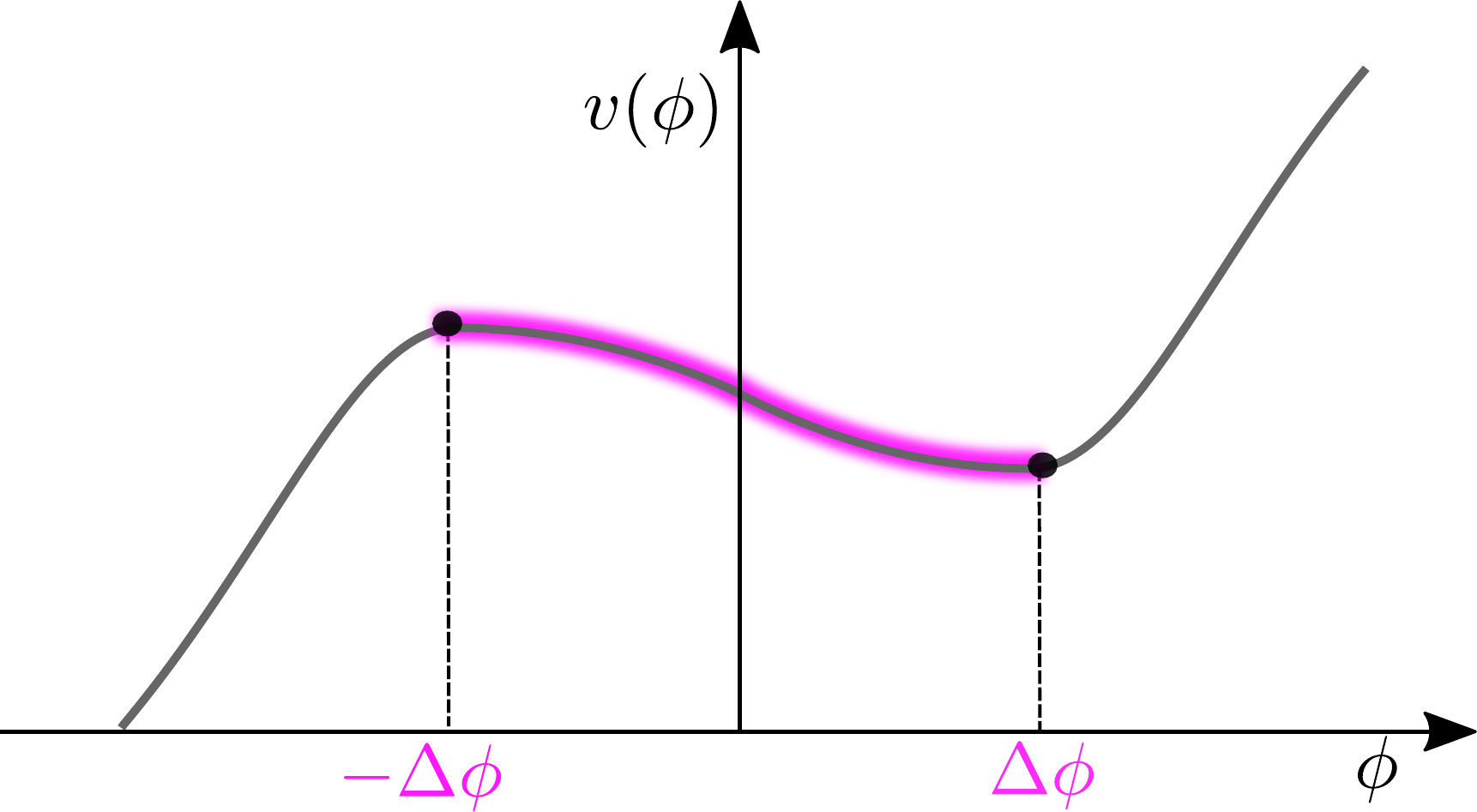}
 \caption{Schematic depiction of the linear potential with negative slope (left panel) and of the piecewise-quadratic potential (right panel) that we consider in this work. We only include the effect of quantum diffusion in the regions highlighted in pink, and assume that the inflaton is mostly driven by the potential gradient elsewhere.}
 \label{fig:potentials}
\end{figure}

Let us now apply this computational programme to a false-vacuum potential. In practice, we assume that slow roll is not violated as the inflaton approaches the local minimum, which is guaranteed~\cite{Pattison:2018bct} if the potential never violates the slow-roll conditions $\epsilon,\vert\eta\vert \ll 1$, where $\epsilon= \Mp^2(v'/v)^2/2$ and $\eta=\Mp^2 (v^{\prime\prime}/ v)$, given that slow roll is a dynamical attractor. Otherwise, a phase of near ultra-slow roll may take place.
We start by considering the toy model displayed in the left panel of \Fig{fig:potentials}, where the potential 
\beq\label{linear}
v(\phi)=v_0 \left(1-\alpha\, \frac{\phi}{\Delta\phi}\right)
\eeq
is linear between the local minimum and the local maximum, which without loss of generality are located at $\phi=\Delta\phi$ and $\phi=-\Delta\phi$ respectively, and where $\alpha$ is a positive parameter.\footnote{The case of a linear potential with a positive slope (\ie negative $\alpha$) was considered in \Refa{Ezquiaga:2019ftu}, and most of the formulas derived below can be checked to be consistent with analytical continuations of those derived in \Refa{Ezquiaga:2019ftu}, up to small differences in the notations employed, \ie $\alpha=-\alpha_{\text{\cite{Ezquiaga:2019ftu}}}\Delta\phi/\Mp$, $\Delta\phi={\phiuv}/2$, $x=2(x_{\text{\cite{Ezquiaga:2019ftu}}}-1)+1$ and $a=-1/(2 a_{\text{\cite{Ezquiaga:2019ftu}}})$ (where $x$ and $a$ are introduced below).} We assume that outside this region, the classical drift provides the main contribution to the inflaton motion and that quantum diffusion can be neglected. This is why, in practice, a reflective boundary condition is placed at $\phi=\Delta\phi$ (such that the inflaton cannot climb up the region it comes from), and an absorbing boundary condition is placed at $\phi=-\Delta\phi$ (such that the inflaton necessarily falls away from the false vacuum once it has reached the local maximum). These translate into
\bea
\label{eq:chi:BC}
\chi(t,-\Delta\phi)=1\quad\text{and}\quad\frac{\partial}{\partial\phi}\chi(t,\Delta\phi)=0\, .
\eea

\begin{figure}[t]
\centering 
\includegraphics[width=.49\textwidth]{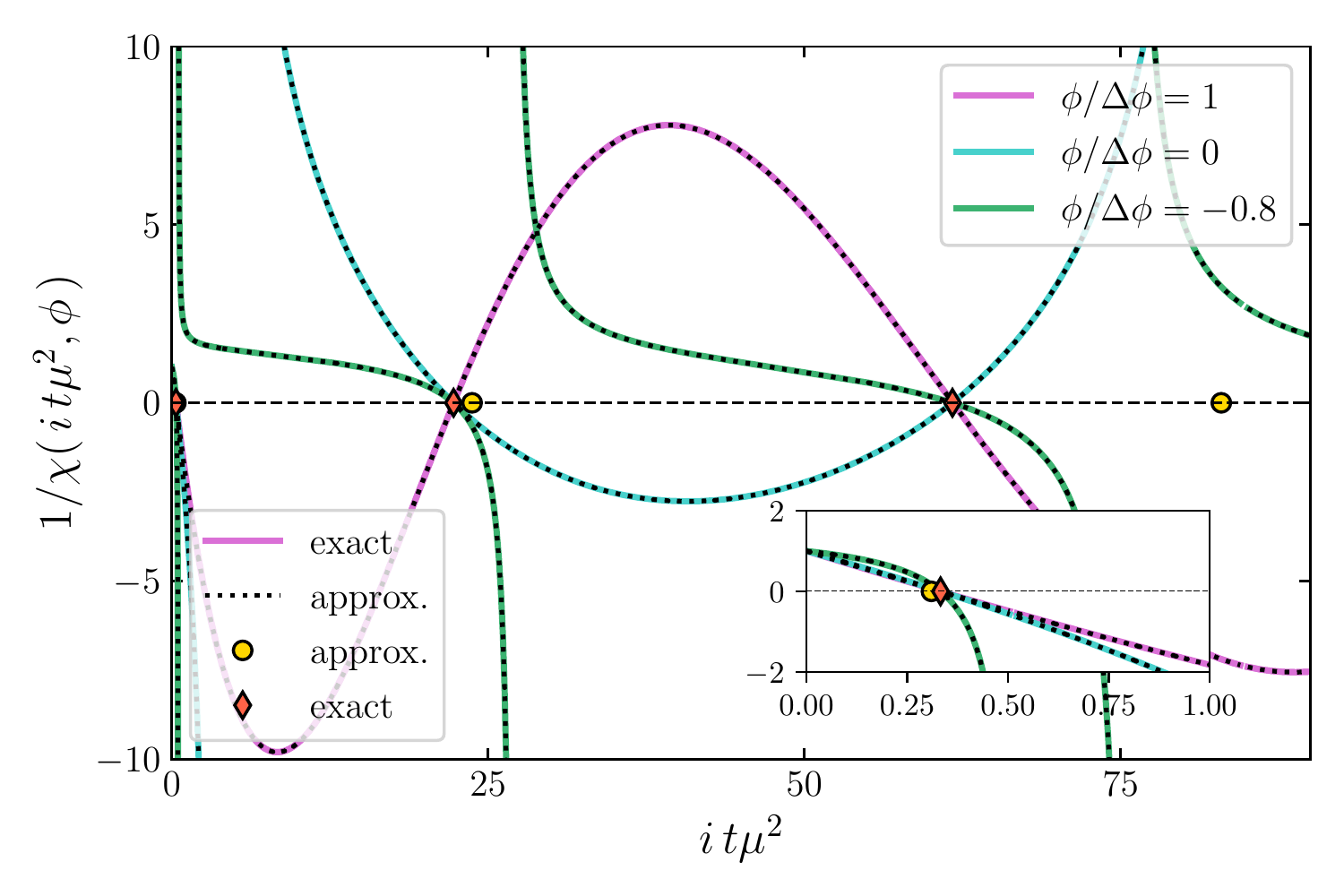}
\includegraphics[width=.49\textwidth]{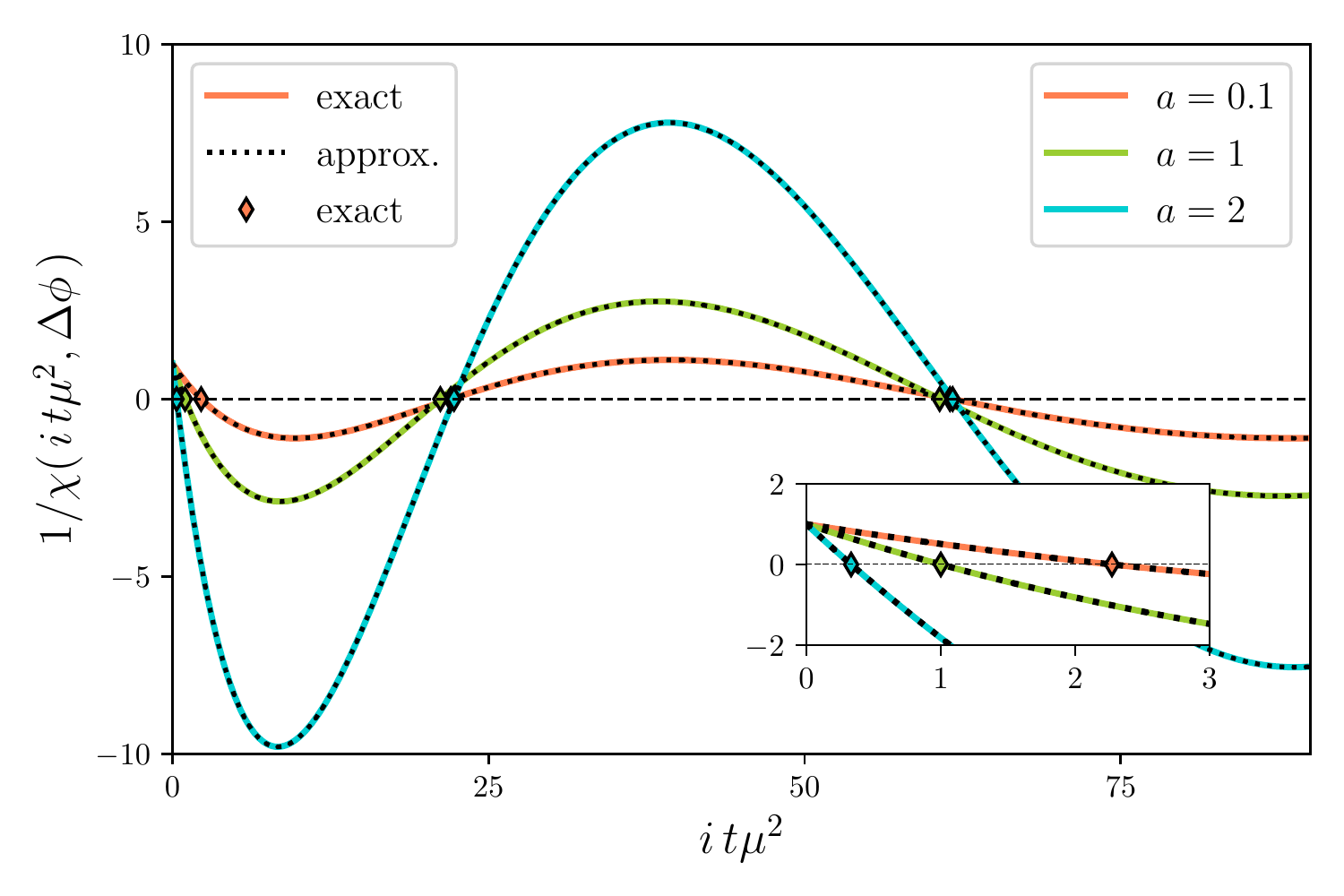}
 \caption{Inverse characteristic function for the linear potential with negative slope, for $a=2$ and a few values of $\phi$ in the left panel, and for $\phi=\Delta\phi$ and a few values of $a$ in the right panel. The solid coloured lines correspond to a numerical solution of the differential equation~\eqref{eq:adjoint:FP:chi} with $v_0=10^{-10}$, while the black dotted lines show the approximated solution~\eqref{chi_lin}, which provides an excellent fit.  On the left panel, one can check that the poles (\ie the values of $t$ where $1/\chi$ intersects $0$) are independent of $\phi$, as expected. In both panels, diamonds denote the pole exact locations, while the approximations derived in the text (namely \Eqs{Lambda_NW}, \eqref{Lambda_WW}, \eqref{Lambda0_WW} or~\eqref{Lambda0a1}  depending on the case under consideration) are displayed with circles in the left panel.
 The agreement is always very good except in the \deep-well ($a>1$) regime at large $n$ (which already occurs at $n=2$ for $a=2$ in the left panel). }
 \label{fig:invchiLin}
\end{figure}

As mentioned above, the slow-roll conditions must be satisfied, which here implies that $\alpha\ll \Delta\phi/\Mp$ (there is no condition coming from $\eta$ since the second derivative of the potential is assumed to identically vanish in this toy model). Furthermore, as we will see below, for the typical number of \efolds spent in the false vacuum to be smaller than $\sim 50$, the height of the potential barrier $\Delta v= v(-\Delta\phi)-v(\Delta\phi)$ has to be much smaller than $v_0$, which also implies that $\alpha\ll 1$. In this regime, the differential equation~\eqref{eq:adjoint:FP:chi} for the characteristic function can be approximated as
\bea
\frac{\partial^2}{\partial\phi^2}\chi(t,\phi)+\frac{\alpha}{v_0\, \Delta\phi}\frac{\partial}{\partial\phi}\chi(t,\phi)+i\frac{t}{v_0\, \Mp^2} \chi(t,\phi)=0\, .
\eea
With the boundary conditions~\eqref{eq:chi:BC}, the solution is given by
\bea\label{chi_lin}
\chi(z,x)=\frac{e^{-\frac{a}{2}(1+x)}\left\lbrace z \cos{\left[\frac{1}{2}(x-1)z\right]}+a \sin{\left[\frac{1}{2}(x-1)z\right]}\right\rbrace}{z \cos{z}-a \sin{z}}\,,
\eea
where we have introduced
\bea
\label{eq:var:redef}
z=\sqrt{i\mu^2 t-a^2}
\quad\text{and}\quad
x=\frac{\phi}{\Delta\phi}\, ,
\eea
together with the parameters
\beq\label{parameters}
\mu^2=\frac{4\, \Delta\phi^2}{v_0\, \Mp^2}
\quad\text{and}\quad
a=\frac{\alpha}{v_0}\,,
\eeq
in terms of which the following discussion takes a clearer form. The parameter $\mu$ may be seen as measuring the amplitude of quantum-diffusion effects, while $a$ parametrises the slope of the potential ($a=0$ for a flat potential), hence the depth of the false vacuum. The approximation~\eqref{chi_lin} is compared with a full numerical solution of the differential equation~\eqref{eq:adjoint:FP:chi} in \Fig{fig:invchiLin}, where one can check that the agreement is indeed excellent.

One may also note that \Eq{chi_lin} is such that $\chi(t=0,\phi)=1$, which from \Eq{chi} implies that the first-passage-time distribution is always normalised to unity. This means that ``infinite inflation'' as discussed in \Refs{Assadullahi:2016gkk, Vennin:2016wnk} does not occur in this setup, and that the field always tunnels out of the false-vacuum state in finite time, which precludes the existence of ever-inflating regions.

\subsubsection{Mean number of \efolds}
\label{subsubsec:constraintsLIN}
\begin{figure}%[t]
     \centering
     \includegraphics[width=0.7\textwidth]{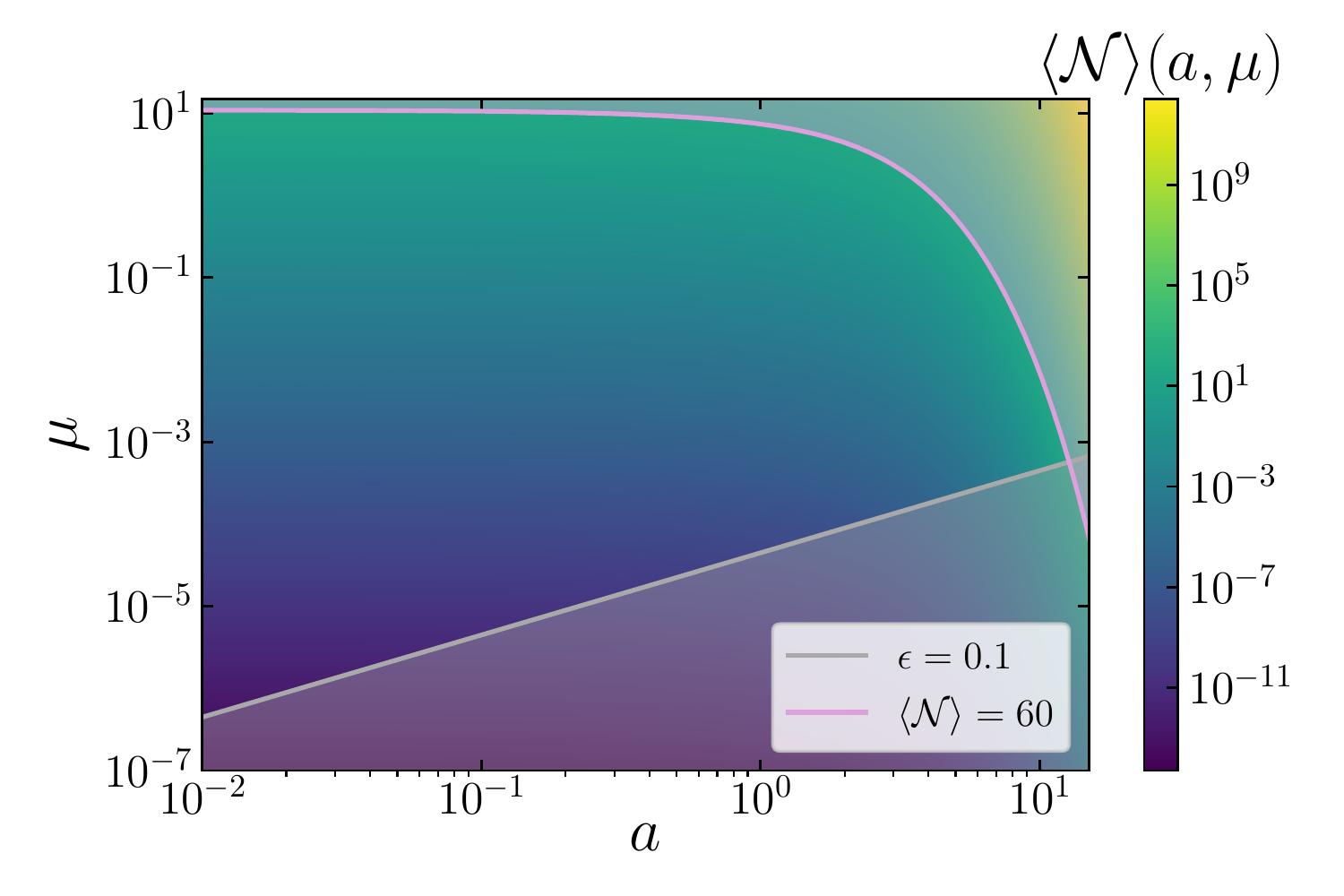}
     \caption{Mean number of \efolds $\mean{\N}$ to escape from the linear false vacuum, starting from the bottom ($x=1$), as obtained from \Eq{N_lin}. The pink line corresponds to $\mean{\N}=60$, above which the model affects CMB and larger scales and should thus be discarded. The grey line corresponds to where the first slow-roll parameter, $\epsilon=2a^2v_0/\mu^2$, is $\epsilon=0.1$. Below that line, the slow-roll approximation does not apply, and nor does our analysis. The remaining region constitutes the relevant region of parameter space on which we focus. Note that $\epsilon$ is computed assuming $v_0=10^{-10}$, which is the largest value allowed by current measurements of the CMB power-spectrum amplitude and upper bound on the tensor-to-scalar ratio~\cite{Planck:2018jri}. If $v_0$ is smaller, $\epsilon$ is smaller (at fixed $a$ and $\mu$), which somewhat relaxes the slow-roll constraint.}
     \label{fig:FigEFLin}
 \end{figure}
In order to better frame the relevant region of parameter space, let us compute the mean number of \efolds spent in the false vacuum. Clearly, it has to be smaller than $\sim 50$ in order for the false vacuum to affect scales that are smaller than those probed in the CMB. Upon Taylor expanding the first relation given in \Eq{chi}, one has
\beq\label{meanN}
\langle \mathcal{N} \rangle(\phi)=-i \left.\frac{\partial \chi(t,\phi)}{\partial t}\right|_{t=0}\,,
\eeq
which together with \Eq{chi_lin} reduces to
\beq\label{N_lin}
\langle \mathcal{N} \rangle(x)=-\frac{\mu^2 }{4 a} \left(x+1\right)+\frac{\mu^2 }{4a^2}\left[e^{2 a}-e^{a\left(1-x\right)}\right]\,.
\eeq
The result is displayed in \Fig{fig:FigEFLin} as a function of $a$ and $\mu$, where the shaded region above the pink line corresponds to where $\langle \N \rangle>60$ and is therefore to be avoided in order to preserve CMB scales.\footnote{Due to the contamination effect highlighted in \Refa{Ando:2020fjm}, the upper bound on $\langle \N \rangle$ might be smaller, hence the exclusion zone in \Fig{fig:FigEFLin} and in similar figures below is conservative.}
 We have also displayed the slow-roll condition derived above, which in terms of $a$ and $\mu$ reads $\mu\gg a\sqrt{2 v_0}$. The shaded region below the grey line corresponds to where the slow-roll parameter $\epsilon$ is larger than $0.1$ and our analysis does not apply. The remaining region encompasses values of $\mu$ and of $a$ that can be as large as a few, but not larger. In particular, the exponentials appearing in \Eq{N_lin} makes the mean number of \efolds strongly dependent on the potential's slope: as the height of the potential barrier increases, it takes exponentially more time for quantum fluctuations to tunnel through it. 

\par

With those restrictions in mind, let us now consider the pole equation, which from \Eq{chi_lin} is given by $z \cos{z}-a \sin{z}=0$. There is an obvious solution $z=0$, but one can check that the numerator of \Eq{chi_lin} also vanishes at $z=0$, and that it is in fact not a pole (except in the singular case $a=1$, which will be discussed separately below). The pole equation can thus be re-written as
\bea
\label{eq:pole:tan}
\frac{z}{\tan z} = a\, .
\eea
This equation is transcendental and the structure of its solutions depends on whether $a\ll 1$ or $a\gg 1$, which will be referred to as the ``\shallow-well'' and the ``\deep-well'' limits respectively in what follows.\footnote{In \Refa{Ezquiaga:2019ftu} the same regimes are called ``narrow'' and ``wide'' in the case of a linear potential with a positive slope.}
 As stressed above, $a$ cannot be much larger than order unity, but we will find that the \deep-well approximation is still reliable for the leading poles even when $a$ is of order one, which makes it useful in that regime. This will also allow us to highlight a special property of false-vacuum potentials, namely the appearance of a new pole as soon as $a>1$. Let us finally note that when evaluating the residues with \Eq{residues:alternative}, one finds
 \beq\label{res_lin}
a_n(x)=\frac{2e^{-\frac{1}{2}a(x+1)}z_n}{\mu^2}\frac{z_n\cos{\left[\frac{1}{2}(x-1)z_n\right]}+a\sin{\left[\frac{1}{2}(x-1)z_n\right]}}{(a-1)\cos{z_n}+z_n\sin{z_n}}\,,
\eeq
 where $z_n=z(t=-i\Lambda_n)$.

\subsubsection{The \shallow-well limit}
If $a\ll 1$, the solutions to \Eq{eq:pole:tan} are such that $\tan z\gg z$ and are therefore close to $z=\pi/2+n\pi$, where $n$ is a non-negative integer. Upon plugging $z=\pi/2+n\pi+\delta z$ in \Eq{eq:pole:tan} and expanding in $\delta z$, one can solve for $\delta z$ and find that the poles are located at $t=-i\Lambda_n$ with
\bea\label{Lambda_NW}
\Lambda_n^{\mathrm{\shallow}}=\frac{1}{\mu^2}\,\left[\pi^2\left(n+\frac{1}{2}\right)^2-2 \,a+\order{a^2} \right] .
\eea
Compared to the flat-potential case ($a=0$, as studied in \Refa{Pattison:2017mbe}), a negative slope ($a>0$) thus makes the tails heavier, while a positive slope ($a<0$) makes the tails lighter (in agreement with \Refa{Ezquiaga:2019ftu}). This is consistent with our previous finding that the typical number of \efolds spent in a false vacuum increases with $a$. The approximation~\eqref{Lambda_NW} can be compared with the full numerical solution of \Fig{fig:invchiLin} in the case where $a=0.1$ (right panel), where one can check that the agreement is indeed excellent.

For the residue, by inserting \Eq{Lambda_NW} into \Eq{res_lin} and further expanding in $a$, one obtains
\bea
\label{a_NW}
a_n^{\mathrm{\shallow}}(x)=&\frac{\pi}{\mu^2}(-1)^n (2 n +1)\cos\left[\frac{\pi}{4}(2 n +1)(x-1)\right]
-\frac{a}{2\mu^2} (-1)^n(x+1)\,\times\\
&\left\{(2n+1)\pi\cos\left[\frac{\pi}{4}(2n+1)(x-1)\right]-2\sin{\left[\frac{\pi}{4}(2n+1)(x-1)\right]}\right\} \\
& +\order{a^2}\, .
\eea
Compared to the flat-well case where $a=0$~\cite{Pattison:2017mbe}, the inclusion of a negative slope $a>0$ thus slightly decreases the leading residue $a_0$, hence the amplitude of the tail (which is nonetheless heavier because of the negative correction to $\Lambda_0$).

\begin{figure}[t]
\centering 
\includegraphics[width=.49\textwidth]{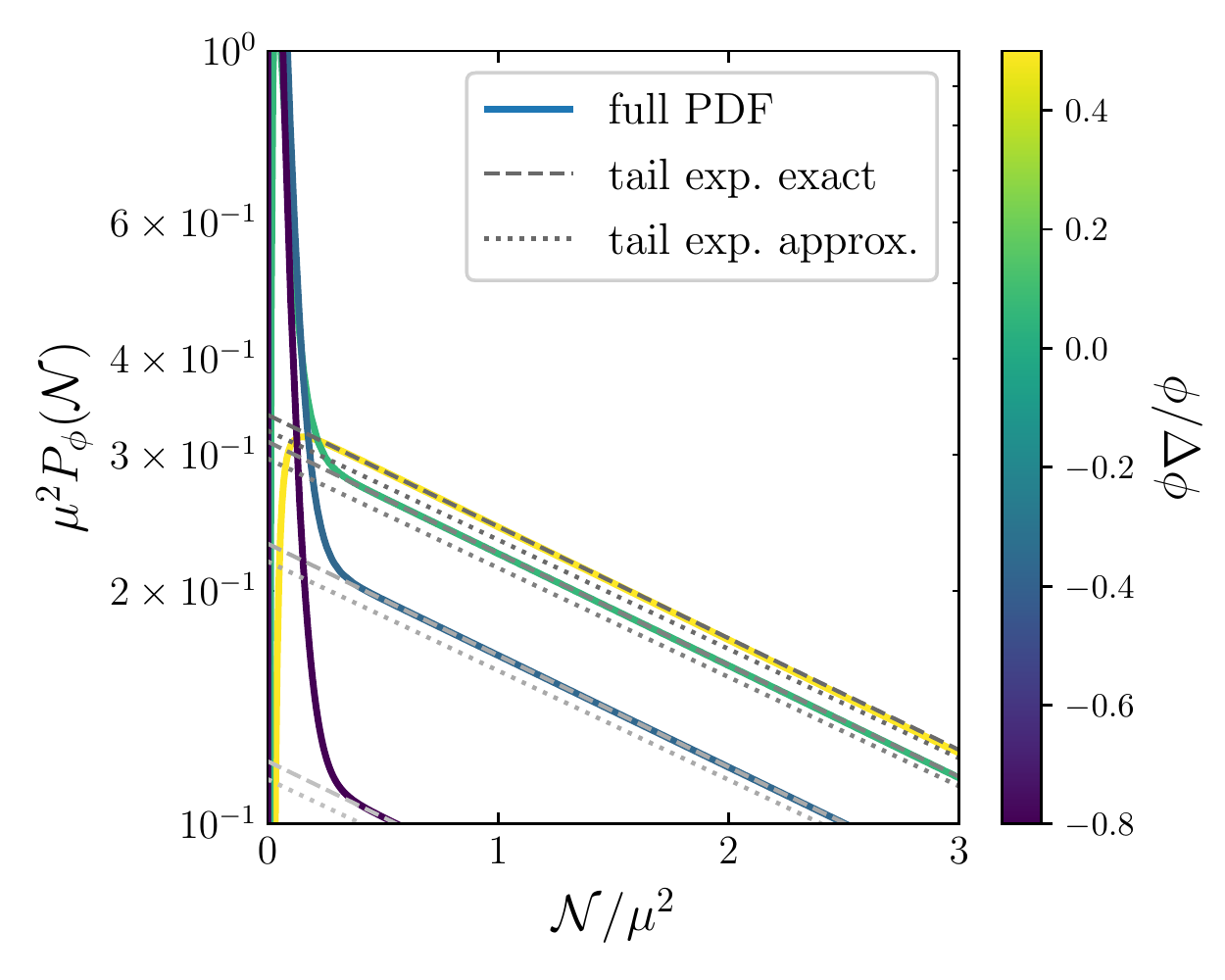}
\includegraphics[width=.49\textwidth]{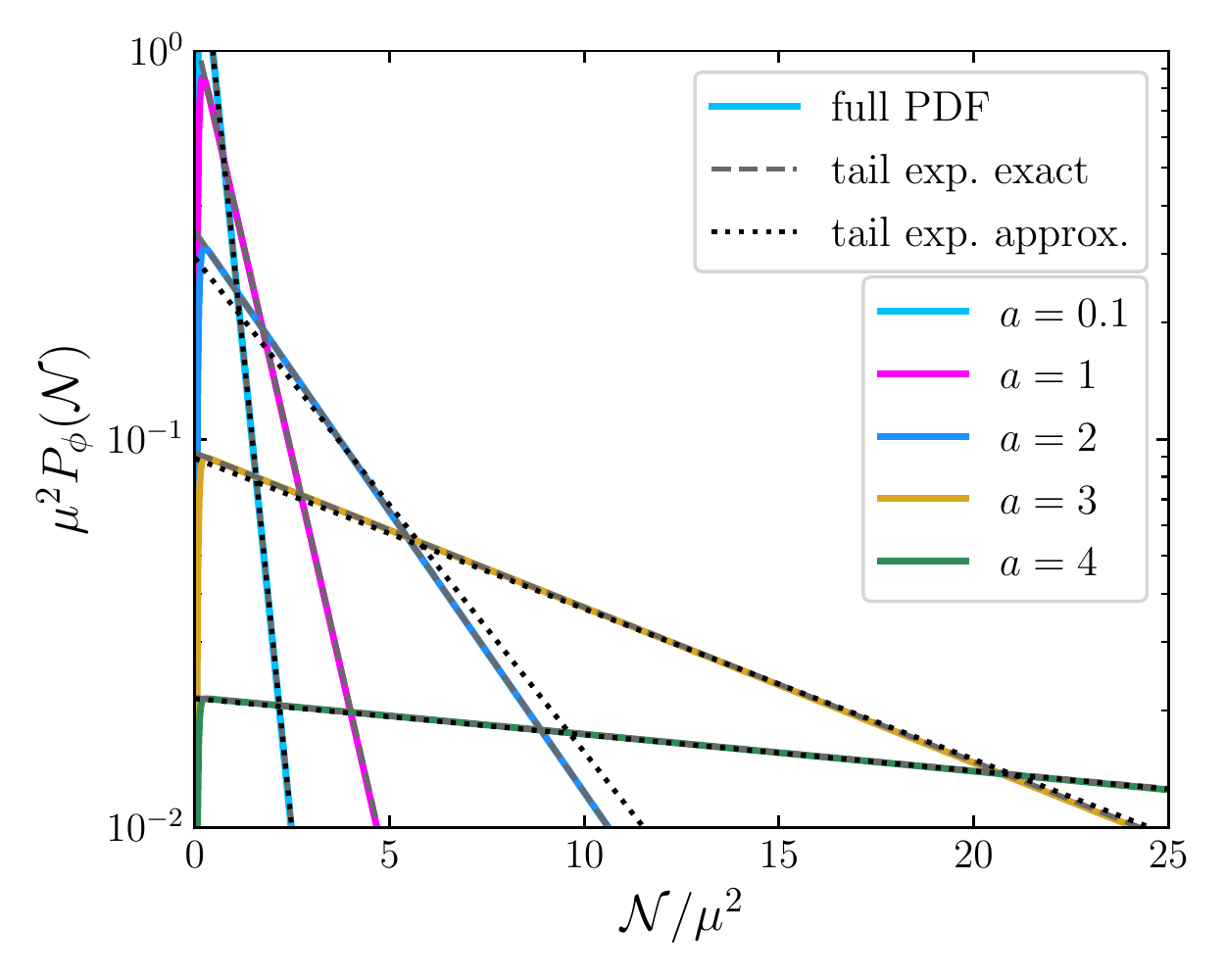}
 \caption{Probability density function of the number of \efolds $\N$ in the linear potential with negative slope~\eqref{linear}, for $a=2$ and a few values of $\phi$ (left panel), and for $\phi=\Delta\phi$ and a few values of $a$ (right panel). The full result, obtained from numerically Fourier transforming the characteristic function (solid coloured lines) is compared with the leading term $n=0$ in the tail expansion~\eqref{general_PDF}, where the first poles and residues have been obtained either numerically (dashed lines) or by the analytical approximations derived in the text (dotted lines, namely \Eqs{Lambda_NW} and~\eqref{a_NW} when $a<1$, \Eqs{Lambda0_WW} and~\eqref{a0} when $a>1$, and \Eqs{Lambda0a1} and~\eqref{a01} when $a=1$). Note that the axes are rescaled by $\mu^2$ such that, due to the self-similarity of the PDF, the result is independent of $\mu$. } 
 \label{fig:PDFLin}
\end{figure}
The PDF can then be obtained from \Eq{general_PDF}. On the tail the leading pole dominates, $P(\N,\phi)\simeq a_0(\phi)\ee^{-\Lambda_0\N}$, so one has
\bea
\label{eq:PDF:linear:shallow:tail}
P^{\mathrm{\shallow}}(\N,\phi=\Delta\phi) \simeq \frac{\pi}{\mu^2}(1-a)\ee^{-\left(\frac{\pi^2}{4}-2a\right)\frac{\N}{\mu^2}}
\eea
where the result is evaluated at the bottom of the false vacuum $\phi=\Delta\phi$. It is displayed for $a=0.1$ with the dotted line in \Fig{fig:PDFLin} (right panel), where it is compared with the full result obtained by numerically Fourier transforming the characteristic function according to \Eq{invFT}, and it is found to fit the tail very well. Close to the maximum of the PDF, the negative slope introduces a correction to the PDF that is parametrically suppressed by $a$ and is thus mostly negligible in the shallow-well regime. On the tail however, the PDF is enhanced by a factor $\sim \ee^{2a\N/\mu^2} \simeq \ee^{a \N /\langle \N \rangle}$, which becomes large for values of $\N$ that are $\sim 1/a$ standard deviations away from the mean.
\subsubsection{The \deep-well limit}
If $a\gg 1$, the solutions to \Eq{eq:pole:tan} are such that $\tan z\ll z$ and are now close to $z=\pi+n\pi$. By expanding \Eq{eq:pole:tan} around that value, one finds
\beq\label{Lambda_WW}
\Lambda_{n+1}^{\mathrm{\deep}}=\frac{a^2}{\mu^2}+\frac{\pi^2}{\mu^2}\,(n+1)^2\left[1+\frac{2 }{a}+\order{\frac{1}{a^2}}\right]\, .
\eeq
Note that this assumes that $z$ is real, \ie that $\Lambda_n\geq a^2/\mu^2$. When $a<1$, one can show that \Eq{eq:pole:tan} does not admit imaginary solutions indeed, but when $a>1$, this is not true anymore: there is one imaginary solution, corresponding to $\Lambda_0$, which is therefore not captured by \Eq{Lambda_WW} (this is why the poles are labeled with the index $n+1$ in \Eq{Lambda_WW}). An approximate expression for this missing pole can be obtained by further expanding $z_0=\sqrt{\Lambda_0 \mu^2-a^2}\simeq i a - i \Lambda_0 \mu^2/(2a) $ in the large-$a$ limit, inserting this formula into \Eq{eq:pole:tan} and solving for $\Lambda_0$. One obtains
\begin{figure}[t]
\centering 
\includegraphics[width=.49\textwidth]{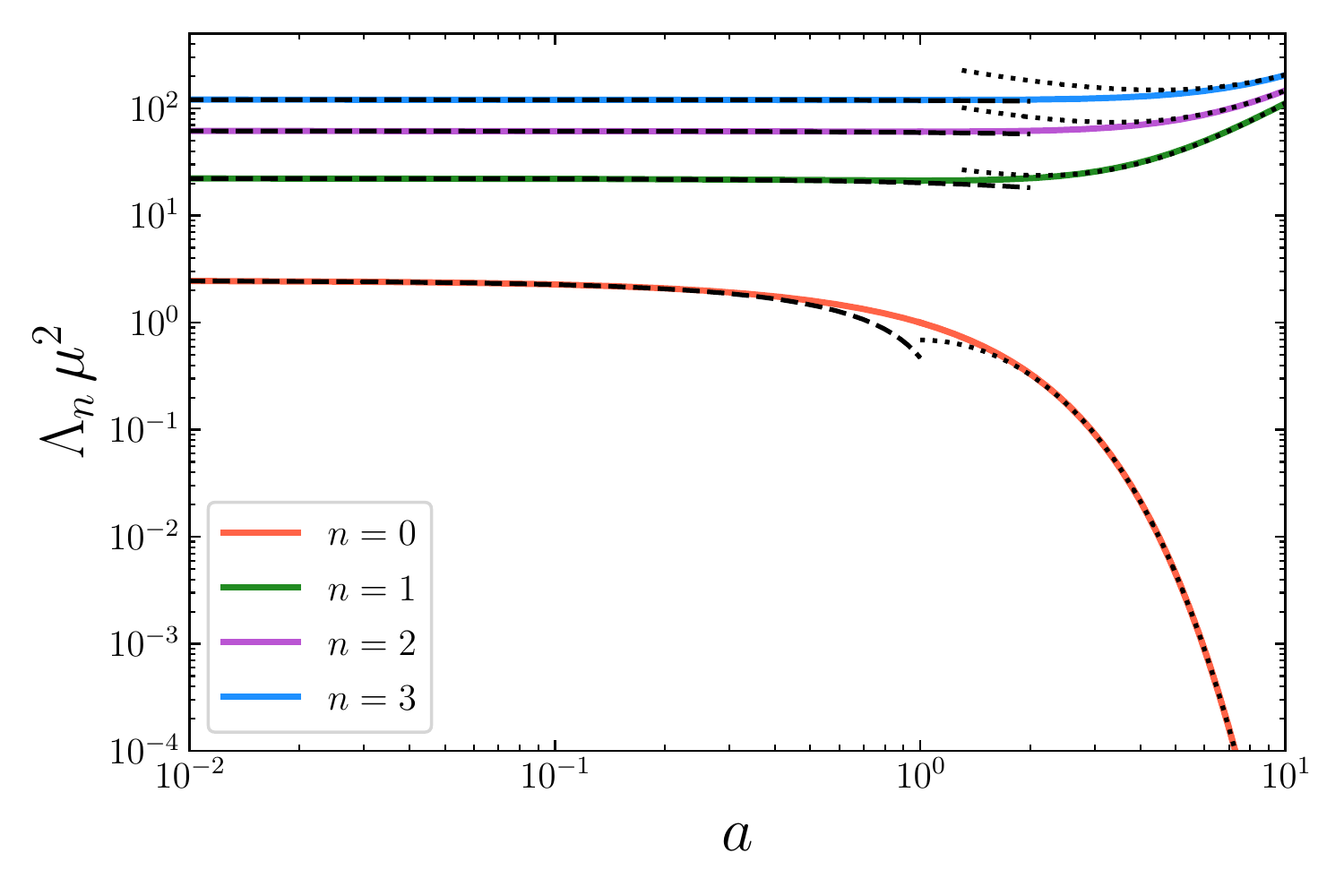}
\includegraphics[width=.49\textwidth]{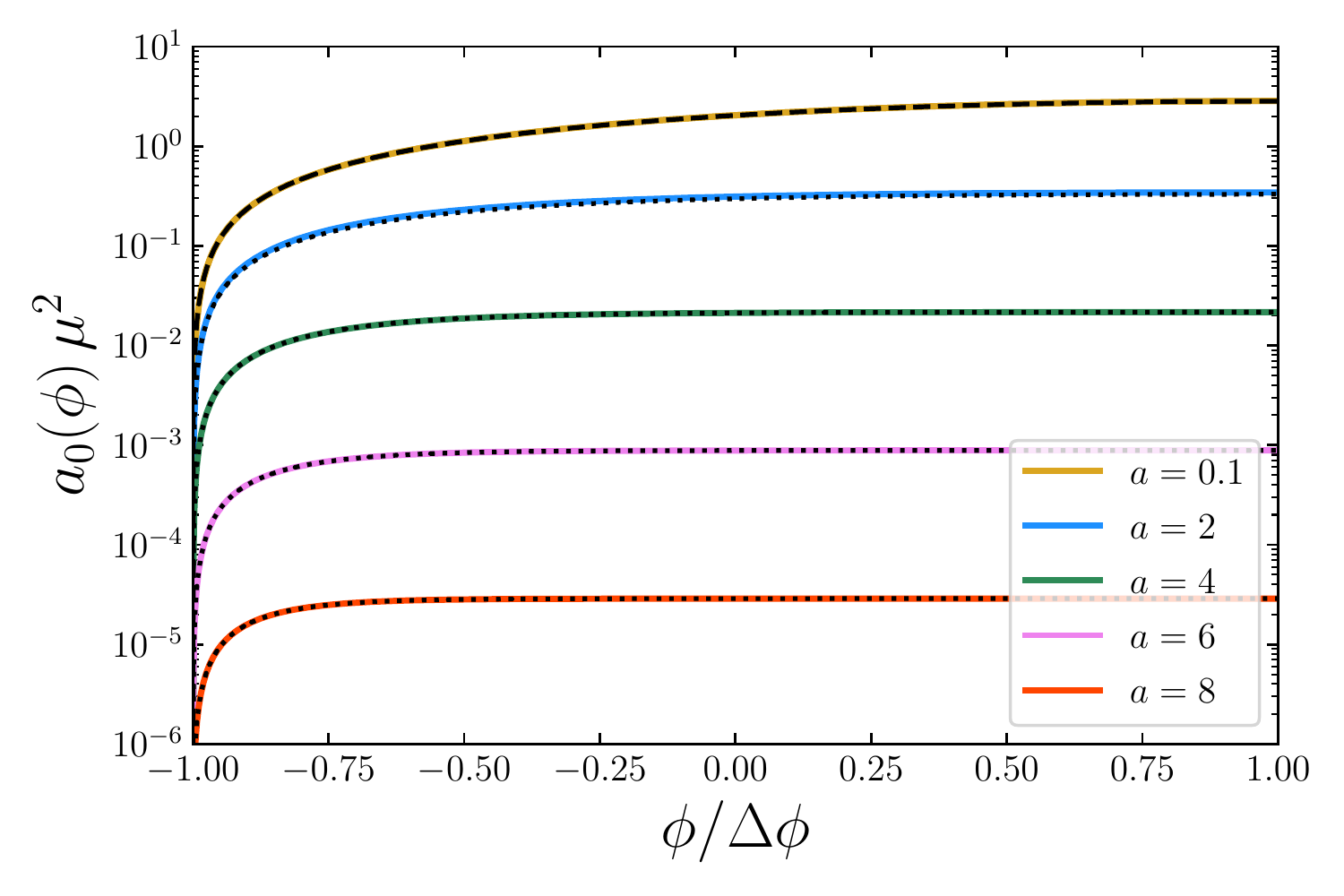}
 \caption{Decay rates in the linear potential with negative slope, for the few first values of $n$ and as a function of $a$ (left panel), and the leading residue for a few values of $a$ and as a function of $\phi$ (right panel), both rescaled by $\mu^2$. Solid lines stand for numerical results obtained from \Eq{chi_lin}, 
dashed lines display the shallow-well approximation (given by~\Eq{Lambda_NW} in the left panel and by~\Eq{a_NW} in the right panel) and dotted lines display the deep-well approximation (given by~\Eqs{Lambda0_WW} and~\eqref{Lambda_WW} in the left panel and by~\Eq{a0} in the right panel).}
 \label{fig:LeadingTermLin}
\end{figure}
\beq\label{Lambda0_WW}
\Lambda_{0}^{\mathrm{\deep}}=\frac{4 a^2 \ee^{-2a}}{\mu^2}
\left[1+2(2a-1)\ee^{-2a}+\order{\ee^{-4a}}\right]\, .
\eeq
By comparing \Eqs{Lambda_WW} and~\eqref{Lambda0_WW}, one can check that, when $a\gg 1$, $\Lambda_0\ll 1$ and $\Lambda_n\gg 1$ for $n\geq 1$. As a consequence, the $0^{\mathrm{th}}$ pole quickly dominates the tail of the PDF.  Let us stress that this additional pole $\Lambda_0$ does not exist in the case of a linear potential with a positive slope, where all poles are given by \Eq{Lambda_WW} (where the sign of $a$ needs to be flipped, see \Refa{Ezquiaga:2019ftu} -- interestingly, that sign flipping does not affect the leading order, which depends on $a^2$). The fact that it is exponentially suppressed with $a$ implies a very heavy tail, which is expected in the \deep-well regime (namely there is a large probability associated with large escaping times from a deep well), and which is also consistent with \Eq{N_lin}. 

The approximations for the poles in both the shallow-well and deep-well regimes are compared to the exact results in the left panel of \Fig{fig:LeadingTermLin}. In particular, one can check that the leading pole $\Lambda_0$ is well reproduced  by \Eq{Lambda0_WW} for $a>1$ and by \Eq{Lambda_NW} for $a\ll 1$. The higher poles can also be compared with the numerical solution of \Fig{fig:invchiLin} for $a=2$, where one can check that \Eq{Lambda_WW} gives a reasonable fit to $\Lambda_1$ but starts failing for $\Lambda_2$ (we have checked that, as $a$ increases, the value of $n$ at which \Eq{Lambda_WW} breaks down increases too, as can be seen from the left panel of \Fig{fig:LeadingTermLin}).

The residues can be obtained by inserting \Eqs{Lambda_WW} and~\eqref{Lambda0_WW} into \Eq{res_lin} and further expanding in $\ee^{-a}$ and $1/a$ respectively, leading to
\bea\label{a0}
a_0^{\mathrm{\deep}}(x) =&  4\frac{a^2}{\mu^2}\ee^{-2a}\left[1-\ee^{-(x+1)a}\right]\\ &
-4 \frac{a^2}{\mu^2} e^{-4 a}    \left\lbrace (x-7) a+4+ \ee^{-(x+1)a} \left[(x+9)a-4\right] \right\rbrace
+\order{\ee^{-6a}}\, ,\\
a_{n+1}^{\mathrm{\deep}}(x)=& \frac{2\pi}{\mu^2} e^{-\frac{1}{2}\,a\,(x+1)} (-1)^{n+1}  (n+1) \sin{\left[\frac{\pi}{2} (n+1)(x-1)\right]} \left[1+\order{\frac{1}{a}}\right] .
\eea
When $x>-1$, at leading order one has $a_0(x)\simeq 4a^2\ee^{-2a}/\mu^2$, which does not depend on $x$. This can be checked in the right panel of \Fig{fig:LeadingTermLin}, where $a_0(x)$ is found to be mildly dependent on $x$ except when it approaches $-1$, and where the above expression is found to provide an excellent approximation. 

As before, the PDF can then be obtained from \Eq{general_PDF}. On the tail, the leading pole strongly dominates as mentioned above, and one has
\bea
\label{eq:PDF:lin:deep}
P^{\mathrm{\deep}}(\N,\phi=\Delta\phi) \simeq 4\frac{a^2}{\mu^2} \ee^{-2a}\ee^{-\frac{4a^2}{\mu^2}\ee^{-2a}\N}\, .
\eea
This can be checked to provide a reliable approximation to the full numerical result in \Fig{fig:PDFLin}. In particular, it works remarkably well even when $a$ is only mildly larger than one. This makes it particularly useful to discuss the case of mildly deep false vacua, \ie the rightmost corner of parameter space as displayed in \Fig{fig:FigEFLin}, where the tail of the PDF is very heavy.

\subsubsection{Singular case $a=1$}
As mentioned above $a=1$ is singular since $z=0$ is a true pole of the characteristic function in that case, so
\beq\label{Lambda0a1}
\Lambda_0^{(a=1)}=\frac{1}{\mu^2}\,,
\eeq
which is an exact solution. By inserting $a=1$ in \Eq{residues} and letting $z\to 0$, this leads to
\bea\label{a01}
a_0(x)^{(a=1)}=\frac{3 }{2 \mu^2} e^{-\frac{1}{2}(x+1)}(x+1)\,,
\eea
which is again exact. It is thus fortunate that in the case $a=1$, the asymptotic tail can be derived exactly, $P^{(a=1)}(\N,\phi) \simeq 3 e^{-\frac{1}{2}(x+1)}(x+1)/(2\mu^2)\ee^{-\N/\mu^2}$. These expressions are confirmed by the numerical analysis performed in the right panels of \Figs{fig:invchiLin} and~\ref{fig:PDFLin}.

%------------
%SUBSECTION: Quadratic potential: two-parabolas approximation
%------------
\subsection{Quadratic piecewise toy model}
\label{subsubsec:QUAD}

Although the linear toy model discussed above yields simple analytical formulas that provide rather straightforward insight, it may be seen as too simplistic to fully describe the physics of a false vacuum. In particular, the potential function being locally extremal at $\phi=\pm\Delta\phi$, its derivative should vanish there, which is not the case in the linear model. A more refined description of a false vacuum may therefore be provided by a piecewise quadratic potential, 
\bea
\label{eq:v:quadratic:piecewise}
v(\phi)=v_0
\begin{cases}
1+\alpha \left[\left(\frac{\phi}{\Delta\phi}-1\right)^2-1\right] \qquad \text{if} \qquad 0 \leq \phi \leq \Delta\phi\, ,\\
\\
1-\alpha \left[\left(\frac{\phi}{\Delta\phi}+1\right)^2-1\right] \qquad \text{if} \qquad -\Delta\phi \leq \phi \leq 0\, .
\end{cases}
\eea
In this expression, the parameters are arranged in such a way that the potential is made of two parabolas with opposite curvatures, matched at the central point of the well $\phi=0$. The potential's derivative vanishes at $\phi=\pm\Delta\phi$, and it is continuous at the matching point $\phi=0$ (where the second derivative flips sign). The height of the potential barrier, $\Delta v=v(-\Delta\phi)-v(\Delta\phi)$, is proportional to $v_0\alpha$, as in the linear potential. Therefore, the parameters $v_0$ and $\alpha$ play similar roles in both models, which will enable a straightforward comparison. Our goal is to determine which of the conclusions drawn for the linear toy model are generic to false vacua in general, and which depend on the details by which it is realised. 
Since the approach is very similar to the one employed in \Sec{subsec:Linear}, some of the technical details will be deferred to \App{app:additional:formulas}.

The potential function~\eqref{eq:v:quadratic:piecewise} is displayed in the right panel of \Fig{fig:potentials}. As for the linear model, it is endowed with a reflective boundary at $\phi=\Delta\phi$ and an absorbing boundary at $\phi=-\Delta\phi$. For the $\epsilon$ slow-roll parameter to be small, one must impose $\alpha\ll\Delta\phi/\Mp$, while $\alpha\ll 1$ is required for the height of the potential barrier to be small compared to $v_0$. Those two conditions were the same in the linear toy model. However, since the second derivative of the potential does not vanish in the quadratic model, there is another condition coming from $\vert\eta\vert\ll 1$, which reads $\alpha\ll (\Delta\phi/\Mp)^2$.\footnote{One may be concerned with the fact that, since $v''$ is discontinuous at the matching point $\phi=0$, the third slow-roll parameter might diverge. However, for the background field trajectory to remain on the slow-roll attractor, only the first two slow-roll parameters need to be small. Moreover, the fact that $v''$ remains finite guarantees that the first three Hubble-flow functions are finite, hence that the Mukhanov-Sasaki frequency is finite  when crossing the matching point.}

The differential equation~\eqref{eq:adjoint:FP:chi} for the characteristic function can be solved separately in the two domains $\phi<0$ and $\phi>0$, where the solution is denoted $\chi_-$ and $\chi_+$ respectively. When $\alpha\ll 1$, \Eq{eq:adjoint:FP:chi} reduces to
\bea
\label{eq:eom:chi:quadratic}
\frac{\partial^2}{\partial\phi^2}\chi_s(t,\phi)+\frac{2\alpha}{v_0\, \Delta\phi}
\left(1-s\frac{\phi}{\Delta\phi}\right)
\frac{\partial}{\partial\phi}\chi_s(t,\phi)+i\frac{t}{v_0\, \Mp^2} \chi_s(t,\phi)=0
\eea
where $s=\pm 1$, and it can be solved as
\bea\label{eq:chi:quadratic:IntegrationConstantsUnfixed}
\chi_s(t,\phi)=c_s \Her \left[\frac{2 z^2}{ s a},\sqrt{s a}\left(x-s\right)\right]
+ d_s \hyp \left[-\frac{z^2}{ s a},\frac{1}{2},sa\left(x-s\right)^2\right] .
\eea
In this expression, $\Her(n,x)$ denotes the Hermite polynomial [sometimes noted $\Her_n(x)$] and $\hyp$ stands for the Kummer confluent hypergeometric function~\cite{NIST:DLMF}. For convenience we have introduced $z=\mu\sqrt{it}/4$, while $x$, $a$ and $\mu$ are still given by \Eqs{eq:var:redef} and~\eqref{parameters} to allow for a direct comparison with the linear case. The parameters $c_-$, $c_+$, $d_-$ and $d_+$ are four integration constants that can be fixed by imposing the boundary conditions~\eqref{eq:chi:BC}  together with the continuity of $\chi$ and of its field derivative, $\partial\chi/\partial\phi$, at the matching point $\phi=0$.\footnote{The continuity of $\chi$ and $\partial\chi/\partial\phi$ can be shown by similar arguments as those usually employed to demonstrate continuity conditions for the wavefunction obeying the Schr\"odinger equation in a piecewise potential: let us rewrite \Eq{eq:eom:chi:quadratic} as  $\chi'' =F(\chi,\chi',\phi,t)$, where a prime denotes derivation with respect to $\phi$ and where  $F$ is a discontinuous though finite function at $\phi=0$. By integrating this relation over $\phi$ around the matching point, one finds $\chi'(t,\delta)-\chi'(t,-\delta)=\int_{-\delta}^\delta F[\chi(t,\phi),\chi'(t,\phi),\phi,t] \dd\phi$, which vanishes when $\delta\to 0$ since $F$ is finite. This proves the continuity of $\chi'$, hence of $\chi$.} This leads to an expression for $\chi$ that we do not reproduce here since it is not particularly insightful, but which is given in \App{app:additional:formulas:chi}, see \Eqs{chi_pos} and~\eqref{chi_neg}.

One can check that it guarantees that $\chi_s(t=0,\phi)=1$, \ie that the field always escapes from the false-vacuum state in finite time, see the remark made below \Eq{parameters}.

\subsubsection{Mean number of \efolds}
\label{subsec:NefQUAD}
\begin{figure}[t]~\label{fig:FigEFQuad}
     \centering
     \includegraphics[width=0.7\textwidth]{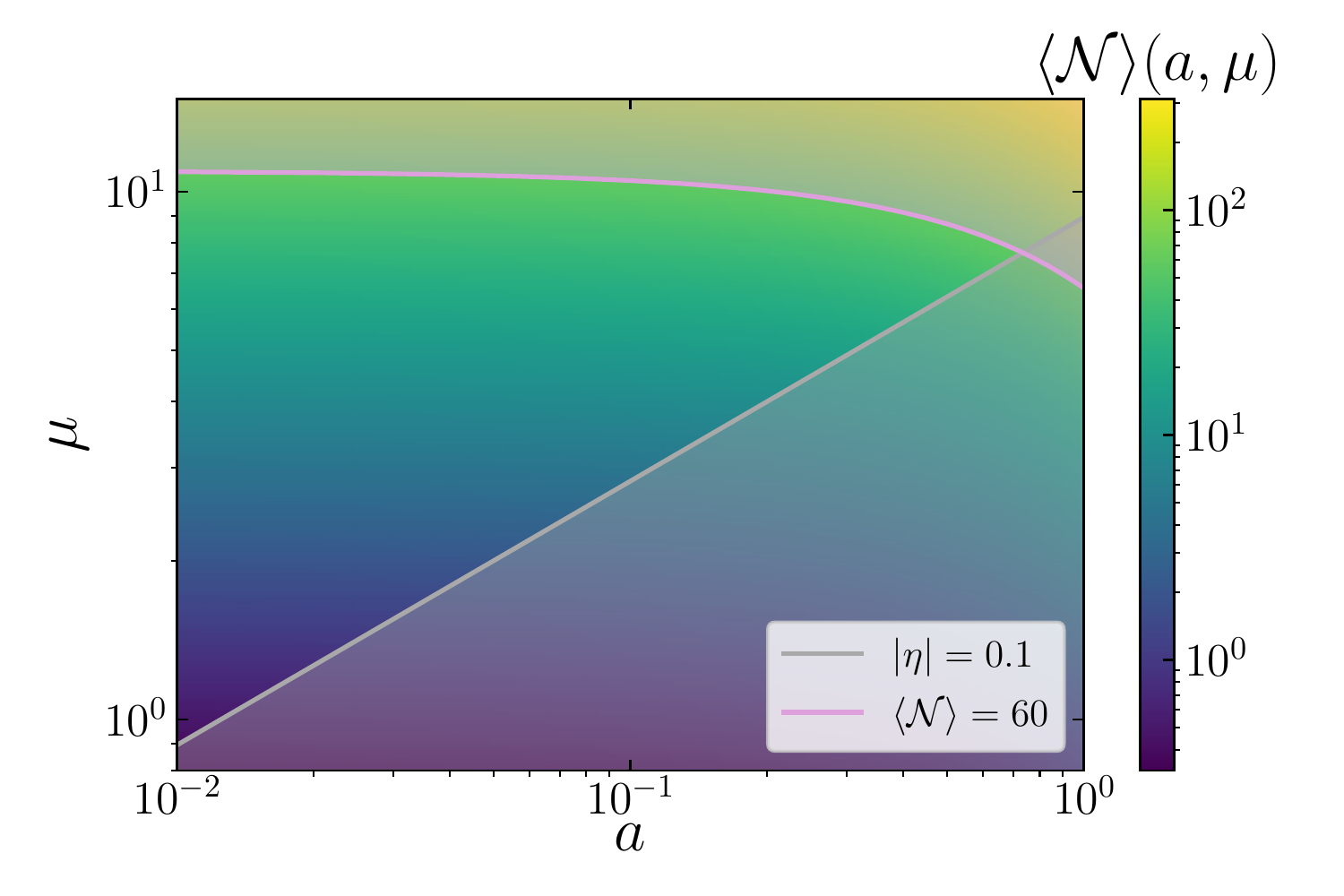}
     \caption{Mean number of \efolds $\mean{\N}$ required to exit the false vacuum in the quadratic piecewise model, starting from $\phi=\Delta\phi$, as given by \Eq{Nplus} and as a function of the parameters $a$ and $\mu$. The pink line corresponds to the contour $\mean\N=60$ so the region located above that line is excluded by CMB observations. The grey line corresponds to $|\eta|=0.1$ and the region located below that line is inconsistent with the use of the slow-roll approximation (the condition coming from $\epsilon$ is less stringent hence it is not displayed and as in \Fig{fig:FigEFLin} we have used $v_0=10^{-10}$ to compute the slow-roll parameters). The remaining, unshaded region thus corresponds to where our analysis applies. }
     \label{fig:FigEFQuad}
 \end{figure}
As for the linear potential, let us now identify the parameters of the model for which the typical time spent in the false vacuum does not exceed a few tens of \efolds. The mean number of \efolds can still be evaluated with \Eq{meanN}, and making use of the above expression for the characteristic function one obtains the formulas~\eqref{Nplus} and~\eqref{Nminus} for $\mean\N$. They are displayed in \Fig{fig:FigEFQuad} when starting from the bottom of the false vacuum at $\phi=\Delta\phi$. $\mean\N$ still features an exponential dependence on the parameter $a$.

Given that $\Delta\phi/\Mp=\sqrt{v_0}\mu/2$, and since $v_0\lesssim 10^{-10}$ while $\mu$ is at most of order $10$, the width of the false vacuum is necessarily sub-Planckian, $\Delta\phi\ll \Mp$. This implies that $|\eta|\gg\epsilon$, hence $\eta$ provides the most stringent slow-roll constraint. The contour $|\eta|=0.1$ is shown in \Fig{fig:FigEFQuad} in order to frame the parameters for which the slow-roll approximation applies. When comparing with \Fig{fig:FigEFLin}, one can see that the structure of parameter space is very similar, the main difference being that values of $a\gtrsim 1$ are now excluded.\footnote{As mentioned before, the position of the contour lines of the slow-roll parameters in the plane $(a,\mu)$ depends on the value of $v_0$, but even when decreasing $v_0$, $a=1$ is never found to be allowed.} The reason for this difference is that, in the quadratic model, the slow-roll conditions involve  $\eta$, hence they are more restrictive. In the linear case one simply sets $\eta=0$ by construction, and this allows for the existence of a ``deep-well'' ($a>1$) regime. The present discussion suggests that such a regime may simply be absent in more realistic models. This is why, below, we only consider the ``shallow-well'' limit.

\begin{figure}[t]~\label{fig:FigInvChiQUAD}
     \centering
     \includegraphics[width=0.7\textwidth]{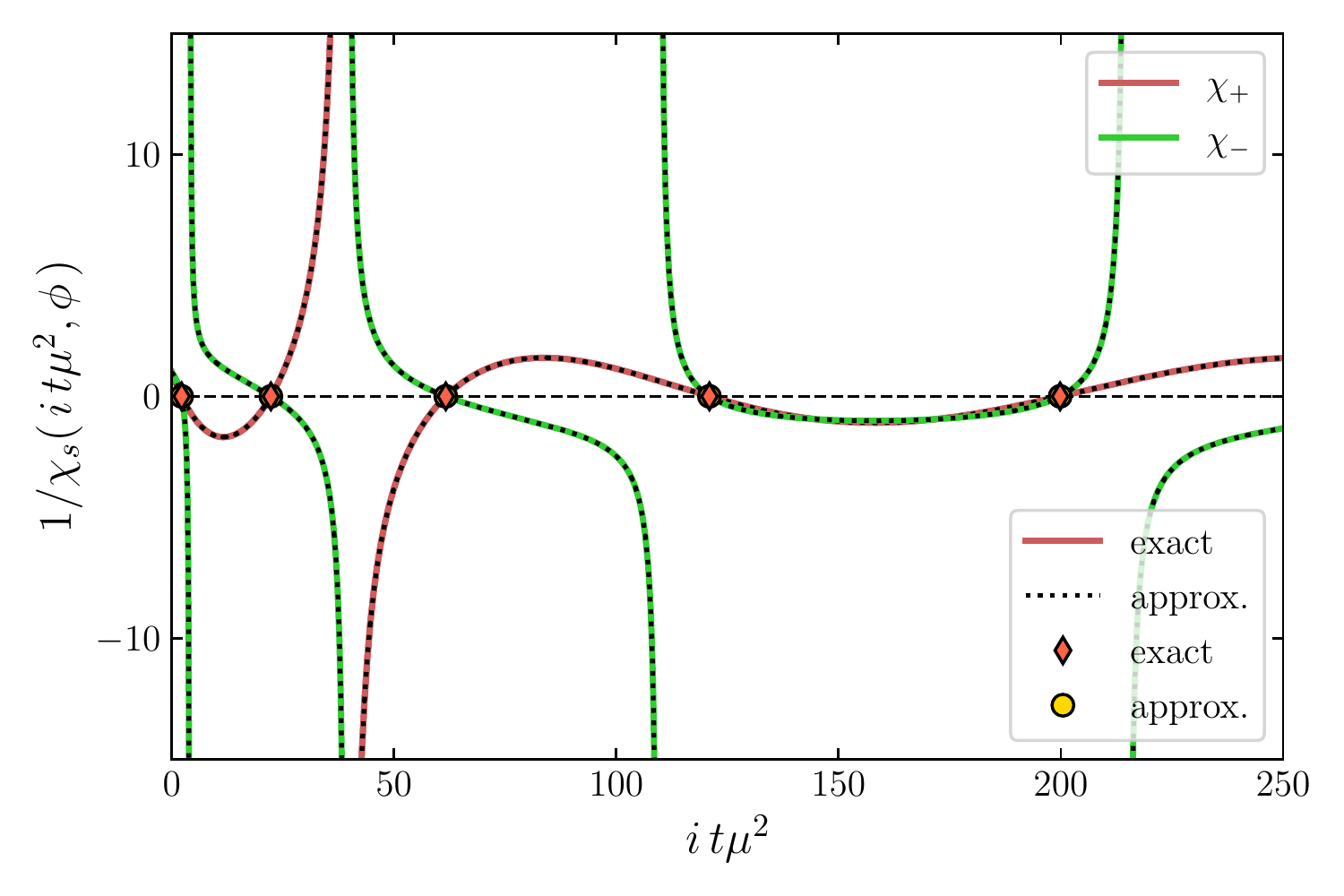}
     \caption{Inverse characteristic function for the quadratic piecewise potential for $a=0.1$ and $v_0=10^{-10}$, $\phi/\Delta\phi=-0.5$ (green line) and $\phi/\Delta\phi=0.5$ (red line). The solid coloured lines correspond to a numerical solution of the differential equation~\eqref{eq:adjoint:FP:chi}, while the black dotted lines show the approximate solution~\eqref{chi_pos}-\eqref{chi_neg}, which provides an excellent fit. The poles (\ie the values of $t$ where $1/\chi$ intersects $0$) are independent of $\phi$, as expected. Diamonds denote the pole exact locations, while the approximation derived in the text, namely \Eq{Lambda_small}, is displayed with circles. The agreement is excellent.}
     \label{fig:FigInvChiQUAD}
 \end{figure}
Before turning to the shallow-well limit, let us note that the poles of the characteristic function are such that the denominators of \Eqs{chi_pos} and~\eqref{chi_neg} vanish, \ie such that
\bea
\label{poleEq}
&\left(1-\frac{z^2}{a^2}\right)\hyp\left(-\frac{z^2}{a}, \frac{1}{2},a\right)^2
+\left(1+2\frac{z^2}{a}\right)^2\hyp\left(-\frac{z^2}{a}, \frac{3}{2},a\right)^2\\ & \qquad\qquad\qquad\qquad\qquad
-2 \left(1+2\frac{z^2}{a}\right)\hyp\left(-\frac{z^2}{a}, \frac{1}{2},a\right)\hyp\left(-\frac{z^2}{a}, \frac{3}{2},a\right)=0\,.
\eea
Upon inspection of \Eq{chi_pos}, one may get the wrong impression that an additional set of poles exist in the branch $\phi>0$, when $\hyp(-z^2/a,1/2,a)=0$. However, one can check that the numerator vanishes too when this condition is realised, such that it does not give any new pole. This is consistent with the fact that the location of the poles must be independent of $\phi$, as shown in \Refa{Ezquiaga:2019ftu}. The inverse characteristic function is displayed in \Fig{fig:FigInvChiQUAD}, where the approximation~\eqref{eq:chi:quadratic:IntegrationConstantsUnfixed} is compared with a numerical integration of \Eq{eq:adjoint:FP:chi}. One can check that the agreement is excellent, and that the same poles are obtained in the two branches $\phi<0$ and $\phi>0$, as expected.
\subsubsection{The shallow-well limit}
\label{sec:quadratic:shallow}
\begin{figure}[t]
     \centering
     \includegraphics[width=0.49\textwidth]{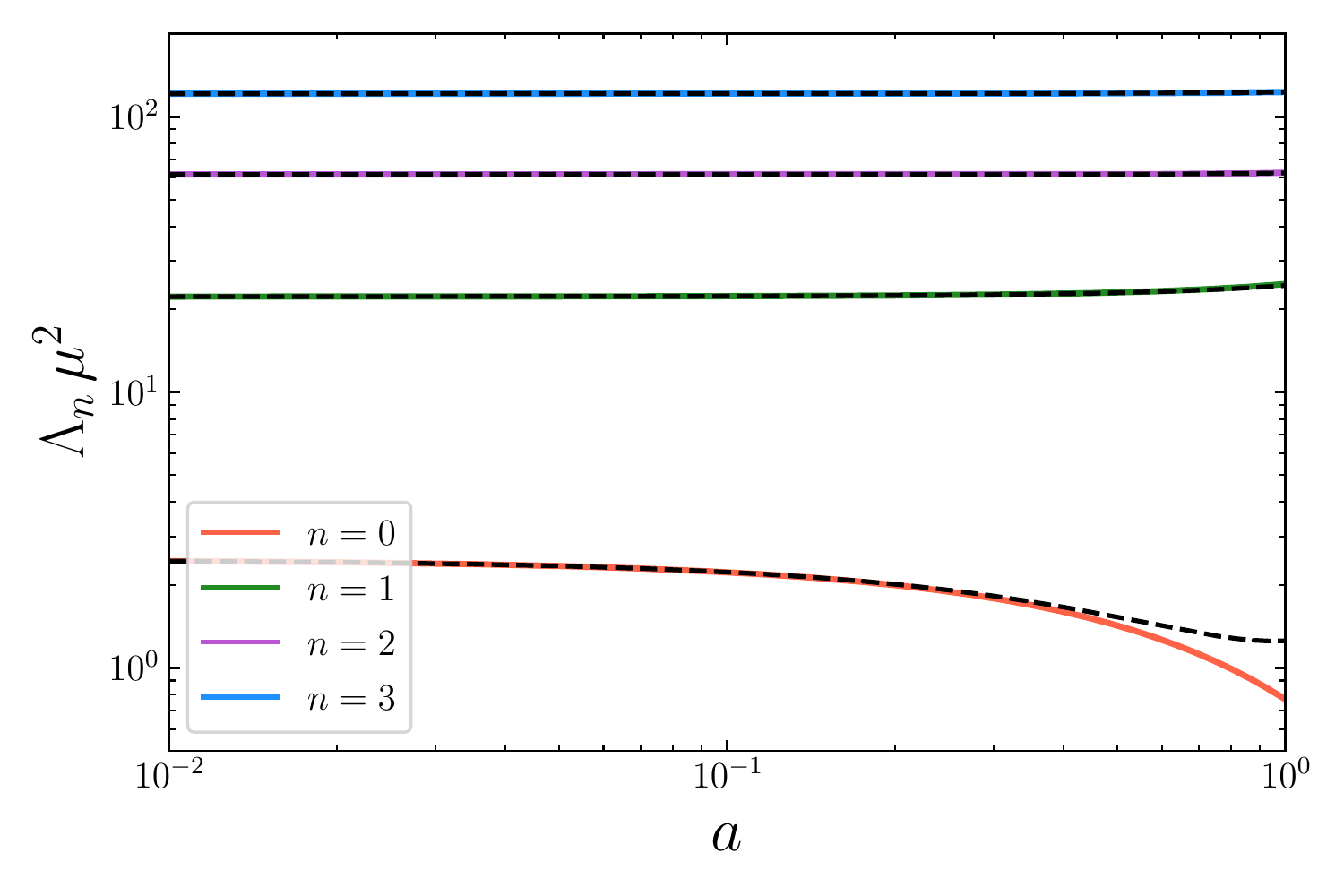}
     \includegraphics[width=.49\textwidth]{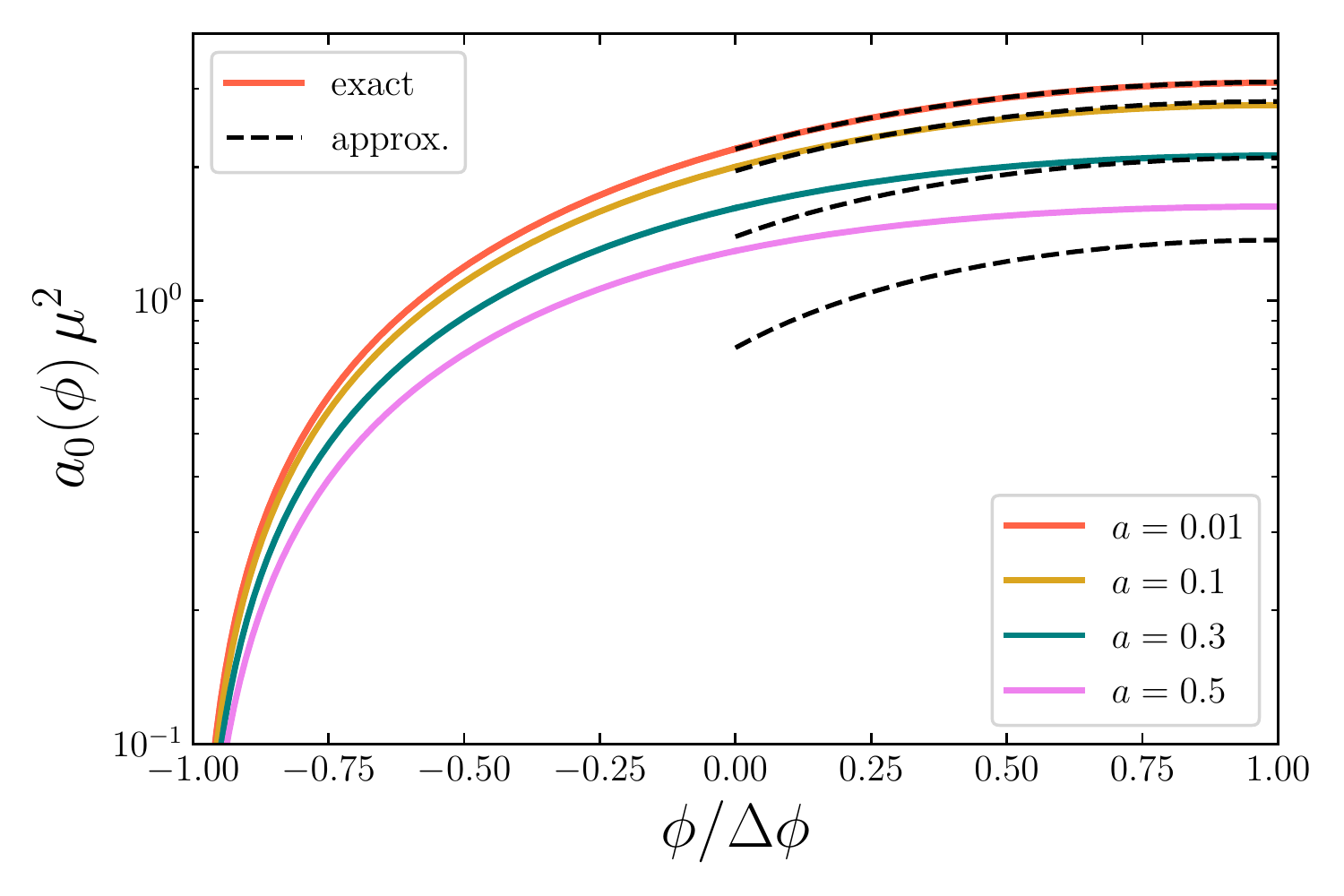}
     \caption{Decay rates in the quadratic piecewise potential, for the first few values of $n$ and as a function of $a$ (left panel), and the leading residue for a few values of $a$ and as a function of $\phi$ (right panel), both rescaled by $\mu^2$. Solid coloured lines stand for full numerical results obtained from \Eq{eq:chi:quadratic:IntegrationConstantsUnfixed}, dashed lines display the shallow-well approximation (given by \Eq{Lambda_small} in the left panel and by \Eq{res_pos_quad:app} in the right panel, only for $\phi > 0$). The agreement is excellent, except when $a$ is order one and at low $n$, as expected.}
\label{fig:polesQUAD}
 \end{figure}

The solutions to the pole equation~\eqref{poleEq} cannot be obtained analytically but as in the linear model, some limits of interest can be studied. Contrary to the linear model however, only the shallow-well limit $a\ll 1$ is relevant, since as explained above $a>1$ cannot be realised without violating the slow-roll conditions. When $a\ll 1$, the first argument of the hypergeometric functions in \Eq{poleEq} becomes large (given that, when $a\to 0$, the roots $z_n$ asymptote the flat-well ($a=0$) result $z_n = \pi(n+1/2)/4$, hence $z_n^2/a=\order{n^2/a}$). Expanding the hypergeometric functions in this regime, in \App{app:pole} we obtain
\bea
\label{Lambda_small}
\Lambda_n=\frac{\pi^2}{\mu^2}\left[\left(n+\frac{1}{2}\right)^2+\frac{4 a^2}{3 \pi^2 }- (-1)^n \frac{ 8 a }{\pi^3 (2 n +1) }\right]+\mathcal{O}\left[(n+1/2)^{-2}\right]\,.
\eea

This expression should be compared with \Eq{Lambda_NW} for the linear-well model. In both cases, the flat-well result is recovered when letting $a\to 0$, and at leading order in $a$, $\Lambda_0$ is decreased by a quantity proportional to $a$, leading to a heavier tail. The approximation~\eqref{Lambda_small} is compared to a numerical solution of the pole equation in \Fig{fig:polesQUAD}, where one can check that the agreement is indeed excellent. One may be surprised to notice that, for higher poles, the approximation is reliable even when $a$ is of order unity, but this is because, as mentioned above, it relies on an expansion on $a/n^2$, which is small even for $a\simeq \order{1}$ when $n$ is large.

The residues can be evaluated by inserting the characteristic function obtained above into \Eq{residues:alternative}. In the shallow-well limit, an expansion in $1/n$ of the corresponding formula is performed in \App{app:residues}, and this leads to
\beq\label{res_pos_quad}
\begin{split}
a_n^{{+}}(x)=&\frac{(-1)^n e^{\frac{1}{2}a[x(x-2)-1]}}{\mu^2}\left\{2 \pi \left(n+\frac{1}{2}\right)\cos{\left[\frac{\pi}{2}\left(n+\frac{1}{2}\right)(x-1)\right]}+\right.\\
&\left.\frac{2}{3}a(x-1)[a(x-2)x-3] \sin{\left[\frac{\pi}{2}\left(n+\frac{1}{2}\right)(x-1)\right]}+\mathcal{O}(n+1/2)^{-1}\right\}
\end{split}
\eeq
when $x>0$, where a formula valid at order $\mathcal{O}(n+1/2)^{-2}$ is also derived in \Eq{res_pos_quad:app}.

This expression should be compared to its counterpart in the linear model, namely \Eq{a_NW}. In both cases, when $a\to 0$ one recovers the flat-well limit, and the first correction controlled by $a$ leads to a decrease of the leading amplitude $a_0$, which can be seen by further expanding \Eq{res_pos_quad} in $a$ once $n=0$ has been fixed. In fact, when starting from the bottom of the false vacuum, $x=1$, both expressions lead to the same result at order $\order{a}$, \ie $a_0(x=1) = (1-a) a_0(x=1)\vert_{a=0}$ in both models. The approximation~\eqref{res_pos_quad:app} is compared to a numerical solution in \Fig{fig:polesQUAD} for $n=0$, which shows that the agreement is excellent for $a \ll 1$ and at the bottom of the false vacuum $(x=1)$ and less accurate for larger values of $a$ when approaching the middle point of the well.
A similar expression can be obtained for $x<0$, but it is found to be less accurate when compared to numerical results. Moreover, in what follows we will be mostly interested in the PDF of the number of \efolds when starting from the bottom of the false vacuum, $x=1$. This makes the expression for $x<0$ of limited interest and this is why we do not reproduce it here.

\begin{figure}[t]
     \centering
     \includegraphics[width=0.49\textwidth]{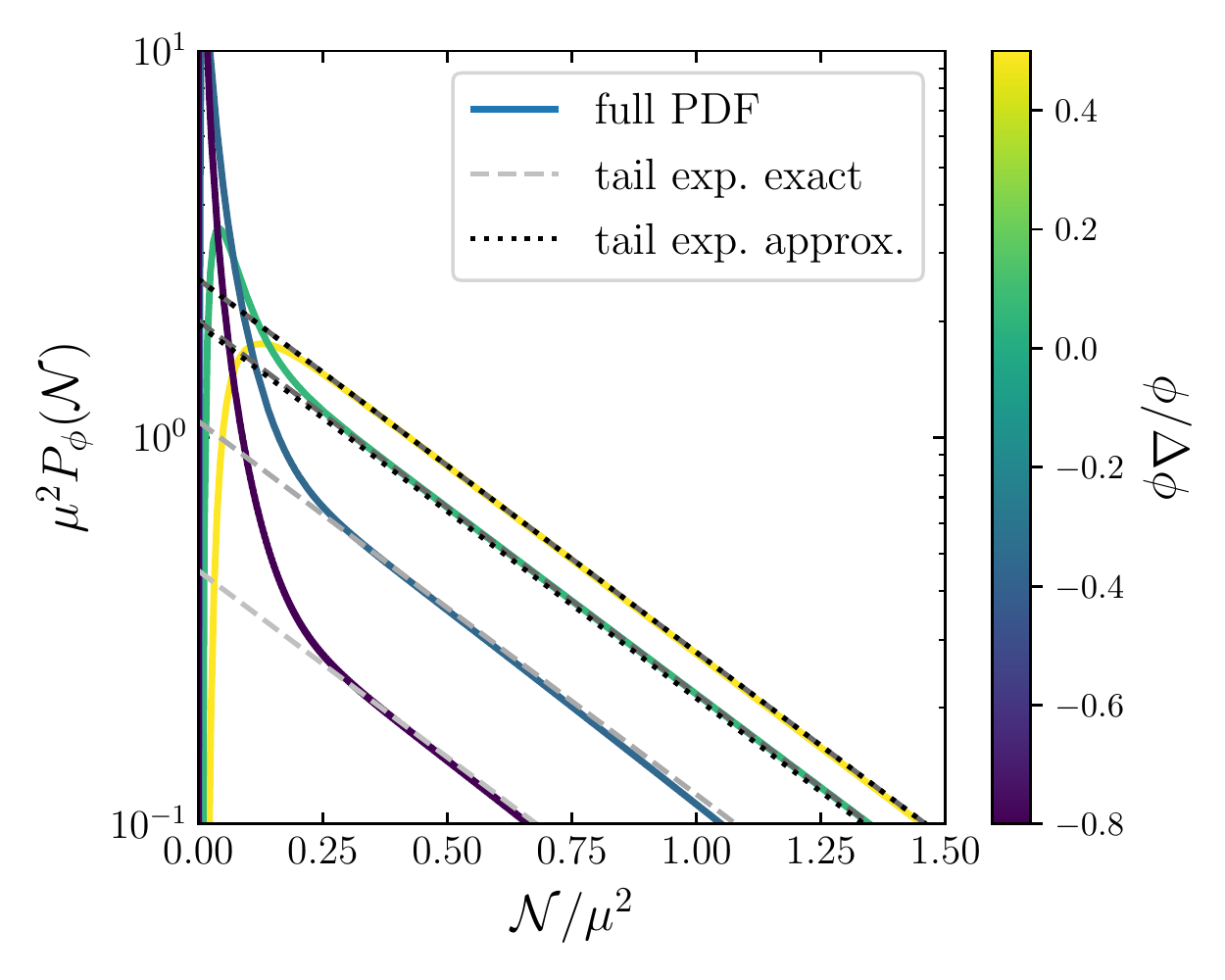}
     \includegraphics[width=0.49\textwidth]{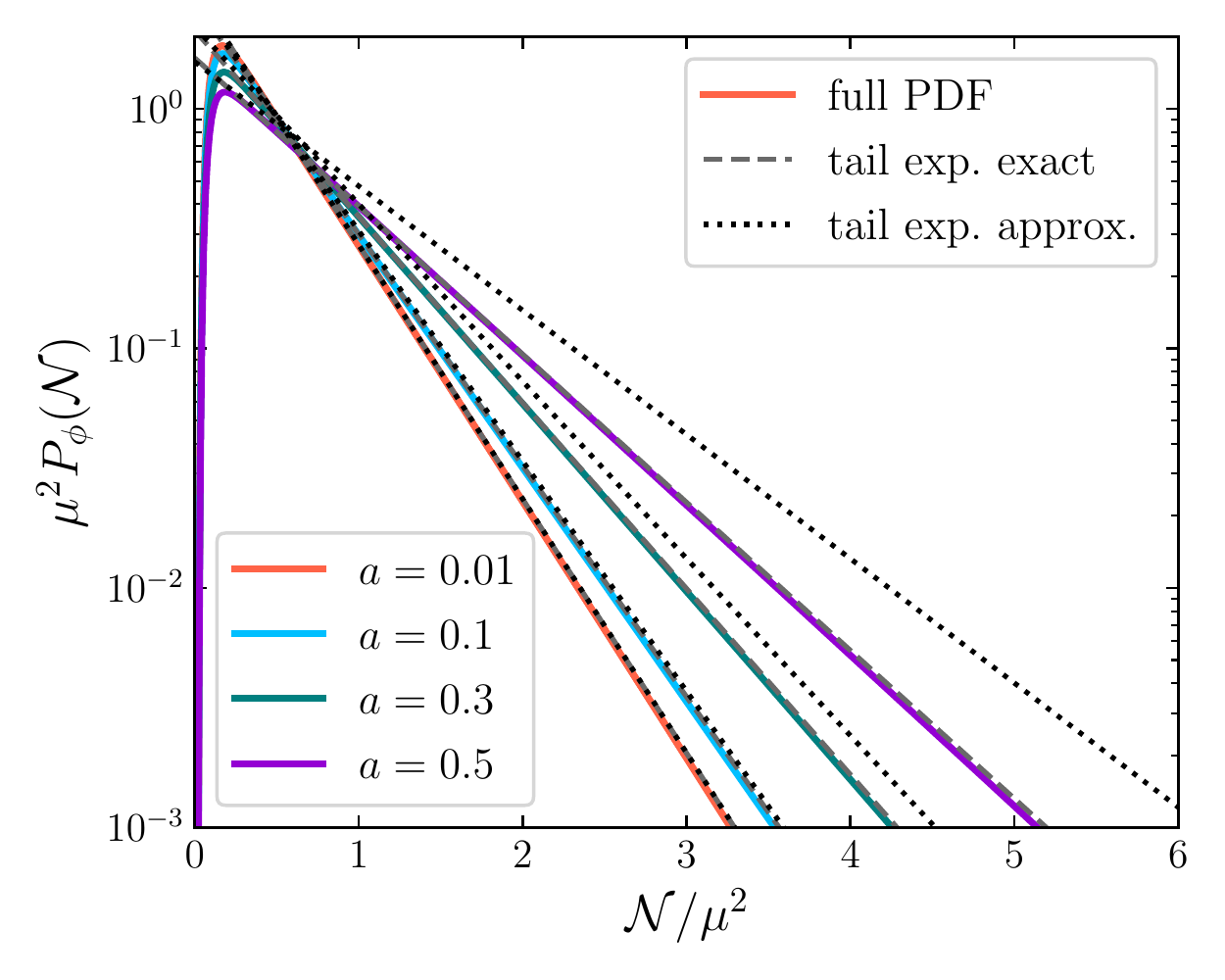}
     \caption{Probability density function of the number of e-folds $\N$ in the piecewise quadratic potential, for $a=0.1$ and a few values of $\phi$ (left panel), and for $\phi=\Delta\phi$ and a few values of $a$ (right panel). The full result, obtained from numerically Fourier transforming the characteristic function (solid coloured lines) is compared with the leading term $n=0$ in the tail expansion~\eqref{general_PDF}, where the first poles and residues have been obtained either numerically (dashed lines) or by the analytical approximations derived in the text (dotted lines, namely \Eqs{Lambda_small} and~\eqref{res_pos_quad:app}, only for $\phi>0$). Note that the axes are rescaled by $\mu^2$ such that, due to the self-similarity of the PDF, the result is independent of $\mu$. }
     \label{fig:pdfquad}
 \end{figure}
Once the poles and the residues have been found, one can compute the PDF of the number of \efolds using \Eq{general_PDF}. The result is displayed in \Fig{fig:pdfquad}, where a full numerical computation is compared with our approximation for the tail, $P(\N)\simeq a_0 \ee^{-\Lambda_0\N}$, which for $x=1$ and further expanding in $a$ reduces to 
\bea
\label{eq:PDF:quadratic:shallow:tail}
P^{\mathrm{\shallow}}(\N,\phi=\Delta\phi) \simeq \frac{\pi}{\mu^2}(1-a)\ee^{-\left(\frac{\pi^2}{4}-\frac{8}{\pi}a\right)\frac{\N}{\mu^2}}\, .
\eea

One can verify in \Fig{fig:pdfquad} that it provides a good fit to the tail, and only starts to be less accurate when $a$ is of order 1.
This expression needs to be compared with \Eq{eq:PDF:linear:shallow:tail} for the linear model, and is found to be very similar, hence the same conclusions apply to both models in the shallow-well regime.

%--------
%PBHs
%--------
\section{Implications for primordial black holes}
\label{sec:PBHs}
\begin{figure}[t]
\centering 
\includegraphics[width=.49\textwidth]{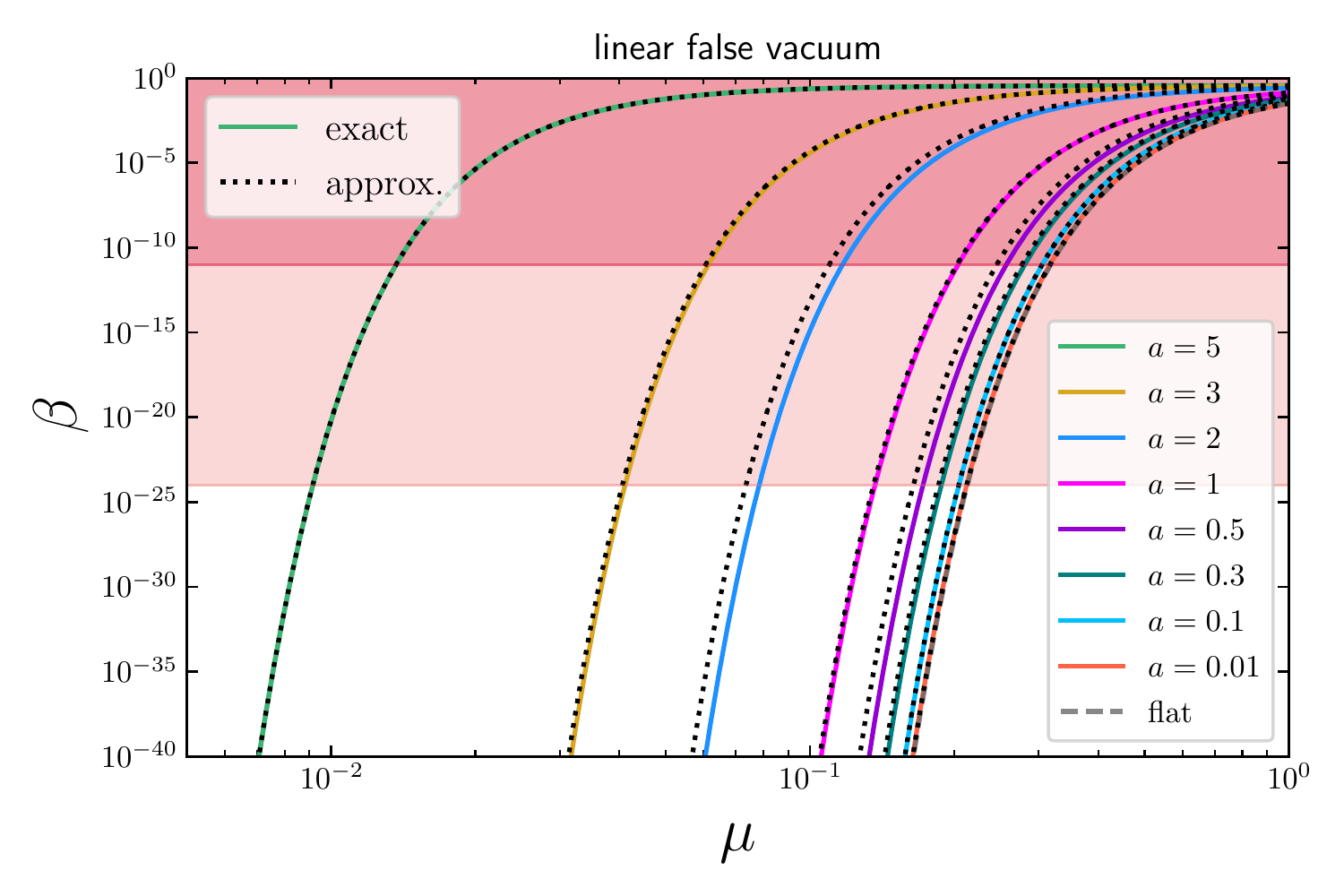}
\includegraphics[width=.49\textwidth]{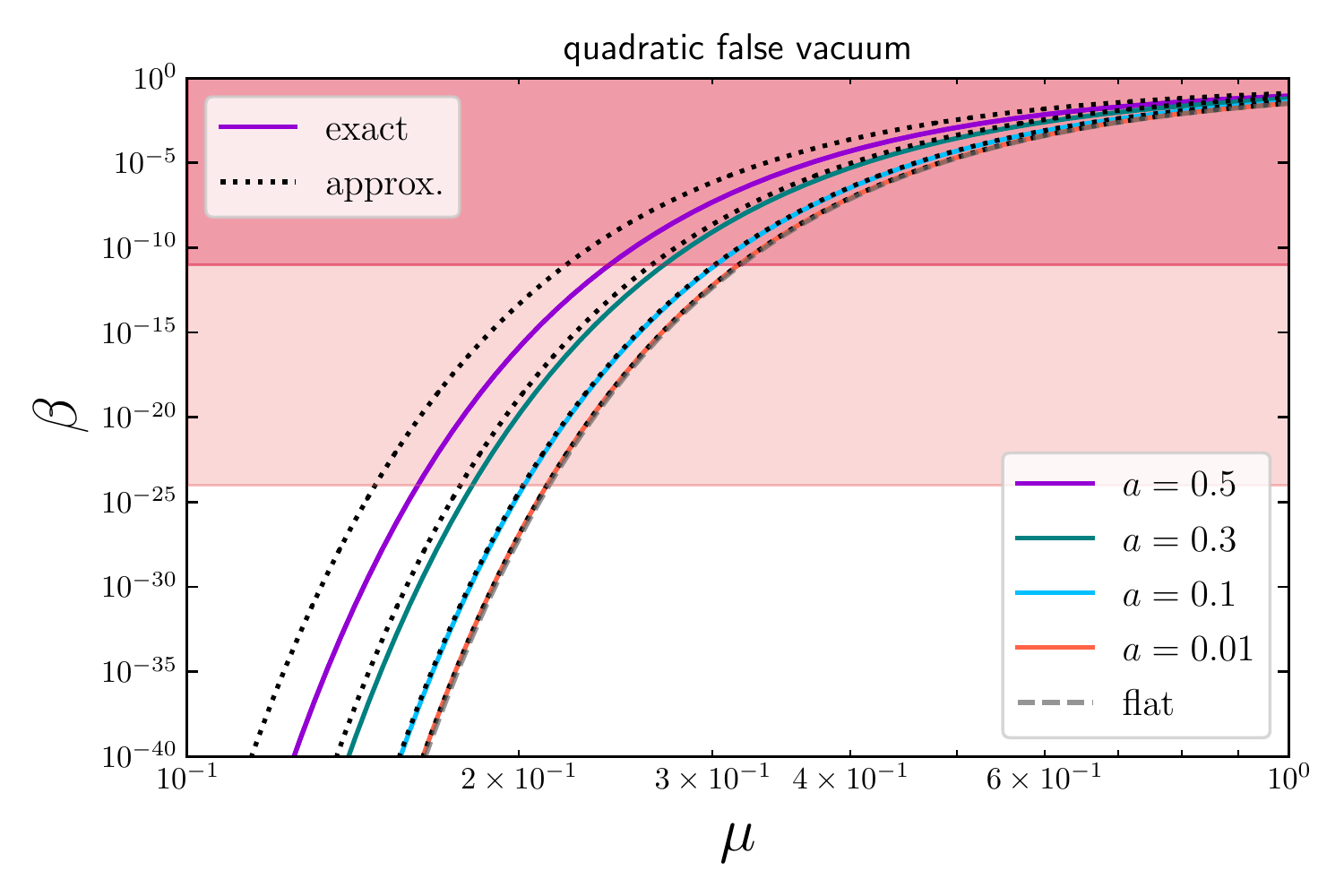}
 \caption{Typical mass fraction $\beta$ of primordial black holes for the linear potential with negative slope (left panel) and for the quadratic piecewise potential (right panel) as a function of $\mu$ for a few values of $a$, assuming $\zetac=1$. We compare the full numerical result (solid line) with our analytical approximations (dotted line), namely \Eqs{eq:beta:linear:shallow},~\eqref{eq:beta:quadratic:shallow},~\eqref{eq:beta:linear:aeq1} or~\eqref{eq:beta:linear:deep} depending on the case of interest. Values of $a>1$ are not displayed in the quadratic model since they lead to slow-roll violation at $\mean\N<60$. The dashed gray line stands for the flat-well limit, \ie $a=0$, for which the difference with $a=0.01$ (red curve) can be barely seen. The two shaded regions stand for the two typical observational upper bounds mentioned in the text, namely $\beta<10^{-24}$ (light pink region) and $\beta<10^{-11}$, the applicability of which depends on the mass at which the black hole forms (which can be tuned by shifting the location of the false vacuum along the inflationary potential, hence it may be seen as an additional free parameter of the models).
  }
 \label{fig:betaLINQUAD}
\end{figure}

Let us now study what the above results imply for the abundance of PBHs in false-vacuum models. There are different criteria for the formation of PBHs, the most advanced ones relying on the compaction function~\cite{Shibata:1999zs, Harada:2015yda, Musco:2018rwt} with critical-scaling relations~\cite{Choptuik:1992jv, Niemeyer:1997mt, Musco:2008hv}, the mass fraction being computed via the excursion-set~\cite{Peacock:1990zz, Bower:1991kf, Bond:1990iw} or the peak-theory~\cite{Bardeen:1985tr} approach. Those methods were however developed for Gaussian or quasi-Gaussian fields, and are therefore not directly applicable to the heavy-tailed distributions obtained in \Sec{sec:Minimum}. Although a research program~\cite{Tada:2021zzj, Kitajima:2021fpq} is underway to generalise them to such highly non-Gaussian statistics, it is yet to be completed and this clearly falls outside the scope of our work. This is why, here, we adopt the simplest viewpoint that PBHs form when the curvature perturbation exceeds a threshold $\zeta_\uc$ of order unity~\cite{Zaballa:2006kh, Harada:2013epa}, and that the resulting mass fraction (\ie the fraction of the universe made of PBHs) is of order the Press-Schechter estimate~\cite{Press:1973iz}
\bea
\label{eq:beta}
\beta\sim \int_{\zeta_\uc}^\infty P(\zeta)\dd\zeta\, .
\eea
This formula makes clear that the abundance of PBHs mostly depends on the tail of the relevant distributions.
In principle, the mass at which the black hole forms is related to the scale at which $\zeta$ is coarse-grained, but here we will use \Eq{eq:beta} as a crude estimate only. The reason is that, as we will see, the PBH abundance is so sensitive on the model's parameters that refining the mass-fraction calculation would clearly not alter our main conclusions. In what follows, we thus evaluate the ``typical'' PBH abundance with the proxy\footnote{Here the PDF of $\N$ is evaluated when starting from the bottom of the potential, $\phi=\Delta\phi$. In principle, the details of the mass distribution are encoded in the way the result depends on $\phi$ (although there is no one-to-one relationship between $\phi$ and the mass, and one rather has to convolve first-passage time distributions with backward probabilities, see \Refa{Tada:2021zzj}). However, as shown in \Refs{Pattison:2017mbe, Ezquiaga:2019ftu} the dependence on $\phi$ is usually mild, and in the present work we only aim at assessing the \emph{typical} abundance of PBHs, without reconstructing their detailed mass distribution.}
\bea
\label{eq:beta:proxy}
\beta\sim\int_{\langle \N \rangle +\zeta_\uc}^\infty P(\N,\phi=\Delta\phi)\dd\N \, ,
\eea
where we have made the identification~\eqref{eq:deltaN:zeta}. Various observational constraints can be set on $\beta$ (see \Refs{Kuhnel:2015vtw, Carr:2009jm} for a comprehensive list of constraints and \Refa{Carr:2020gox} for a more recent  update). To summarise, for masses between $10^9$ and $10^{16}$ g, the constraints come primarily from the effects of PBH Hawking evaporation on Big Bang Nucleosynthesis (BBN) and on the extragalactic photon background, and range from $\beta<10^{-24}$ to $\beta< 10^{-17}$. Heavier PBHs, whose masses are comprised between $10^{16}$ and $10^{50}$ g, have still not evaporated, thus the constraints mostly come from their gravitational and astrophysical effects, and range from $\beta < 10^{-11}$ to $\beta < 10^{-5}$. For ultra-light PBHs, \ie black holes with masses lower than $10^9$ g, which Hawking evaporate before BBN, there are no direct observational constraints on their abundance. As pointed out in \Refs{Martin:2019nuw, Papanikolaou:2020qtd, Auclair:2020csm}, such black holes could even dominate the energy content of the primordial universe during a transient phase, and would leave a stochastic background of gravitational waves as (one of) their only imprint.

Let us stress that, since we have found the PDF of $\N$ to feature exponential, rather than Gaussian, tails, the integral appearing in \Eq{eq:beta:proxy} would be vastly under-estimated by using a Gaussian (or quasi-Gaussian) ansatz, and it is crucial to account for non-Gaussianities by means of non-perturbative techniques such as the stochastic $\delta  N$ formalism employed here~\cite{Pattison:2017mbe, Ezquiaga:2019ftu, Panagopoulos:2019ail, Figueroa:2020jkf, Pattison:2021oen, Achucarro:2021pdh, Hooshangi:2021ubn, Ezquiaga:2022qpw}. This has strong consequences not only for the predicted abundance of PBHs, but also for the viability of a given model in terms of the above-mentioned observational constraints. 

By inserting \Eq{general_PDF} into \Eq{eq:beta:proxy}, one obtains 
\bea
\beta\sim\sum_n \frac{a_n(\Delta\phi)}{\Lambda_n} e^{-\Lambda_n \left[\zetac+\mean\N(\Delta\phi)\right]}\, ,
\eea
where $\mean\N$ can be evaluated with \Eq{meanN}, leading to $\mean\N(\phi)=\sum_n a_n(\phi)/\Lambda_n$. The result is displayed in \Fig{fig:betaLINQUAD} for the linear (left panel) and quadratic (right panel) false-vacuum models. The solid lines stand for a full numerical calculation, \ie they are obtained by numerically integrating the PDFs (themselves obtained by numerically Fourier transforming the characteristic function) above the threshold $\mean\N+\zetac$. The dotted lines correspond to the approximations derived in \Sec{sec:Minimum}, and which we now review.

In the linear toy model, in the \shallow-well regime the PDF can be approximated by \Eq{eq:PDF:linear:shallow:tail} on the tail, and the mean number of \efolds is given by \Eq{N_lin}, which reduces to $\mean\N(\Delta\phi)\simeq \mu^2(1+2a/3)/2$ in the limit $a\ll 1$. Inserting those expressions into \Eq{eq:beta:proxy}, one finds
\bea
\label{eq:beta:linear:shallow}
\beta^{\mathrm{linear,\, \shallow}} \simeq \frac{4}{\pi}\left[1+\left(\frac{8}{\pi^2}-\frac{\pi^2}{12}\right)a\right]\ee^{-\frac{\pi^2}{8}-\left(\frac{\pi^2}{4}-2a\right)\frac{\zetac}{\mu^2}}\, .
\eea
This needs to be compared with the \shallow-well limit of the quadratic toy model. Upon expanding \Eq{Nplus} in the regime $a\ll 1$, one finds $\mean\N(\Delta\phi)\simeq\mu^2(1+5a/6)/2$, and together with \Eq{eq:PDF:quadratic:shallow:tail} this leads to 
\bea
\label{eq:beta:quadratic:shallow}
\beta^{\mathrm{quadratic,\, \shallow}} \simeq \frac{4}{\pi}\left[1+\left(\frac{32}{\pi^3}+\frac{4}{\pi}-\frac{5\pi^2}{48}-1\right)a\right]\ee^{-\frac{\pi^2}{8}-\left(\frac{\pi^2}{4}-\frac{8}{\pi}a\right)\frac{\zetac}{\mu^2}}\, .
\eea
When $a=0$, in both cases one recovers the flat-well result, displayed with the dashed grey line in \Fig{fig:betaLINQUAD}, and for $a<1$ one can check in \Fig{fig:betaLINQUAD} that \Eqs{eq:beta:linear:shallow} and~\eqref{eq:beta:quadratic:shallow} provide a good fit to the full result indeed. This shows that PBHs form in the far-tail region of the PDF, which is dominated by the leading pole. Two main comments are in order at this stage. 

First, we note that the result is very similar in the linear and the quadratic models. In \Sec{sec:Minimum}, we had found that the two models differ in the properties of their higher poles, which scale differently with $n$, but here we find that since PBHs form in the region of the tail that is dominated by the leading pole, they are produced with similar abundances in both models. This makes our conclusions for the \shallow regime generic, since they seem independent of the precise way the false vacuum is realised. 

Second, the introduction of a local minimum through the small parameter $a$ leads to a parametrically small modulation of the prefactor in $\beta$, which is negligible, but it also enhances the mass fraction exponentially, $\beta\sim\beta(a=0)\ee^{A a\zetac/\mu^2}$, where $A=2$ in the linear model and $A=8/\pi$ in the quadratic model. In the quadratic model, the slow-roll condition reads $\mu\gg\sqrt{a}$, so this exponential enhancement is negligible too (recall that $\zetac$ is of order one). We thus reach the important conclusion that, in the quadratic false vacuum, the flat-well approximation applies whenever the slow-roll approximation does. 

In contrast, in the linear model, the slow-roll condition reads $\mu\gg a\sqrt{v_0}$, where $v_0$ needs to be smaller than $10^{-10}$. This implies that the exponential enhancement factor may be large even at small values of $a$, and without violating the slow-roll conditions. More precisely, two cases must be distinguished. If $\mu$ is large, then PBHs are over-produced, $\beta\sim 1$, which is excluded except if they have masses smaller than $10^9\ \mathrm{g}$ and Hawking evaporate before BBN as explained above. In that case, the result is only mildly affected  by $a$, and the flat-well approximation may be used. If $\mu$ is small, then PBHs are produced with small initial abundance, such that they might play a relevant astrophysical role later on. In that case, a linear false vacuum with $\mu^2\lesssim a \ll 1$ yields abundances that are substantially different from the flat-well model, and needs therefore to be properly described. This can be clearly seen in \Fig{fig:betaLINQUAD}: when $\mu\sim 0.1$, slight changes in $a$ result in orders-of-magnitude differences in $\beta$.

When $a$ is not a small parameter, as explained in \Sec{subsubsec:QUAD} the quadratic model cannot be described within the slow-roll approximation while keeping $\mean\N$ small enough, and only the linear model should be discussed. If $a=1$, combining \Eqs{N_lin},~\eqref{Lambda0a1} and~\eqref{a01} leads to
\bea
\label{eq:beta:linear:aeq1}
\beta^{\mathrm{linear,\, }a=1} \simeq 3 \ee^{\frac{3-\ee^2}{4}-1}\ee^{-\frac{\zeta_\uc}{\mu^2}}
\simeq 0.37 \ee^{-\frac{\zeta_\uc}{\mu^2}}\, .
\eea
Recall that this expression is exact on the tail, \ie $a_0$ and $\Lambda_0$ have been derived without performing any approximation in the special case $a=1$. 
In the deep-well limit, expanding \Eq{N_lin} in the regime $a\gg 1$ leads to $\mean\N(\Delta\phi)\simeq \mu^2\ee^{2a}/(4a^2)$, and together with \Eq{eq:PDF:lin:deep} one finds
\bea
\label{eq:beta:linear:deep}
\beta^{\mathrm{linear,\, \deep}} \simeq \ee^{-1}\ee^{-\left(2a\ee^{-a}\right)^2\frac{\zetac}{\mu^2}}\, .
\eea
The two expressions above can be checked to provide good fits to the full numerical result in \Fig{fig:betaLINQUAD}. The mass fraction features super-exponential dependence on $a$ in the deep-well regime, which implies that PBHs are over-produced as soon as $a\gtrsim 8$.

\begin{figure}[t]
\centering 
\includegraphics[width=.49\textwidth]{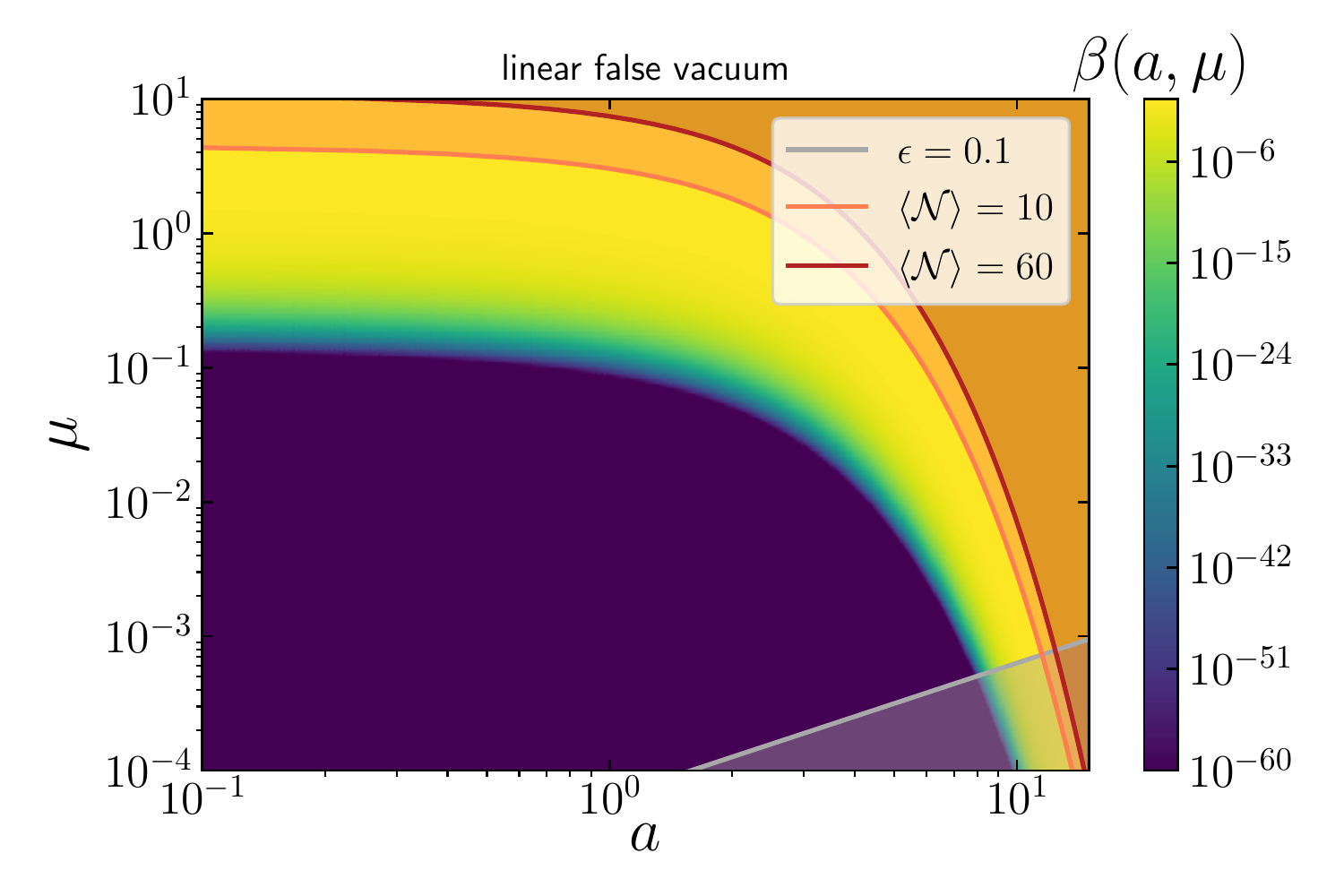}
\includegraphics[width=.49\textwidth]{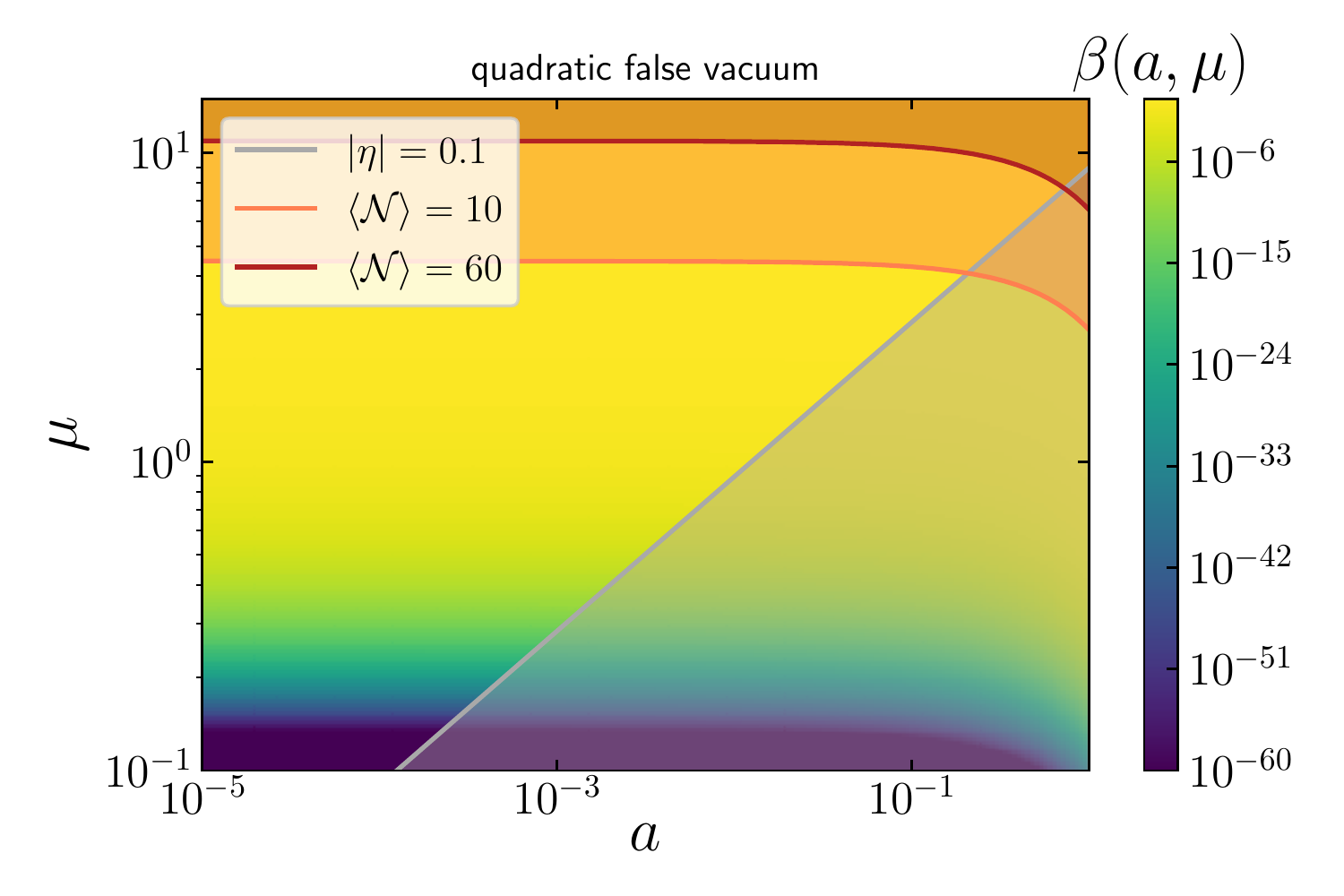}
 \caption{Colour map of the mass fraction $\beta$ of primordial black holes for the linear potential with negative slope (left panel) and for the quadratic piecewise potential (right panel) as a function of $\mu$ and $a$, for $\zetac=1$. In each panel, the grey shaded area corresponds to where slow roll is violated, and the red and pink shaded areas stand for $\mean\N>60$ and $\mean\N>10$ respectively. For convenience, we set the lower limit of the colour bar to $10^{-60}$, such that all the uniform dark violet region corresponds to values of $\beta \leq 10^{-60}$.
 }
 \label{fig:colorbetaLINQUAD}
\end{figure}

\par 

The discussion is summarised in \Fig{fig:colorbetaLINQUAD} where we show the mass fraction in the parameter space $(a,\mu)$. As in \Figs{fig:FigEFLin} and~\ref{fig:FigEFQuad}, we also display contour lines for the mean number of \efolds spent in the false vacuum, and the regions where the slow-roll approximation does not apply (hence our result cannot be trusted) is shaded in grey. 

In the quadratic false vacuum (right panel), imposing that the slow-roll approximation applies and that $\mean\N<60$ leads to PBH abundances that are always well-captured by the flat-well limit ($a=0$)~\cite{Pattison:2017mbe}. The relevant parameter in that case is $\mu$: if $\mu\ll 1$, a tiny amount of PBHs is produced; if $\mu\lesssim 1$, PBHs are produced with sizeable abundances, which makes them of potential astrophysical interest, while if $\mu\gtrsim 1$ they are overproduced, which is excluded if they form with masses larger than $10^9\ \mathrm{g}$, and which can lead to a transient PBH-dominated universe otherwise (large values of $\mu$ are excluded by the condition $\mean\N<60$).

The same considerations apply to the linear false vacuum when $a\ll \mu^2$, but the fact that slow-roll conditions are less stringent in this model leads to the existence of two additional regimes. When $\mu^2\ll a\ll 1$ (which implies that $\mu$ is small), one obtains large deviations from the flat-well reference case, while still being in the shallow-well domain. In this regime, one may obtain PBHs with initial abundances that make them of astrophysical interest, and with mass distributions that carry a non-trivial imprint of the false-vacuum profile and which cannot be simply described by flat-well models. When $a$ is of order one, there is a small corner in parameter space where the slow-roll approximation applies and $\mean\N<60$, and which is well described by the deep-well regime. There, PBHs are more massively produced.

%--------
%CONCLUSION
%--------
\section{Discussion and conclusion}
\label{sec:Conclusion}
\begin{figure}[t]
\centering 
\includegraphics[width=.99\textwidth]{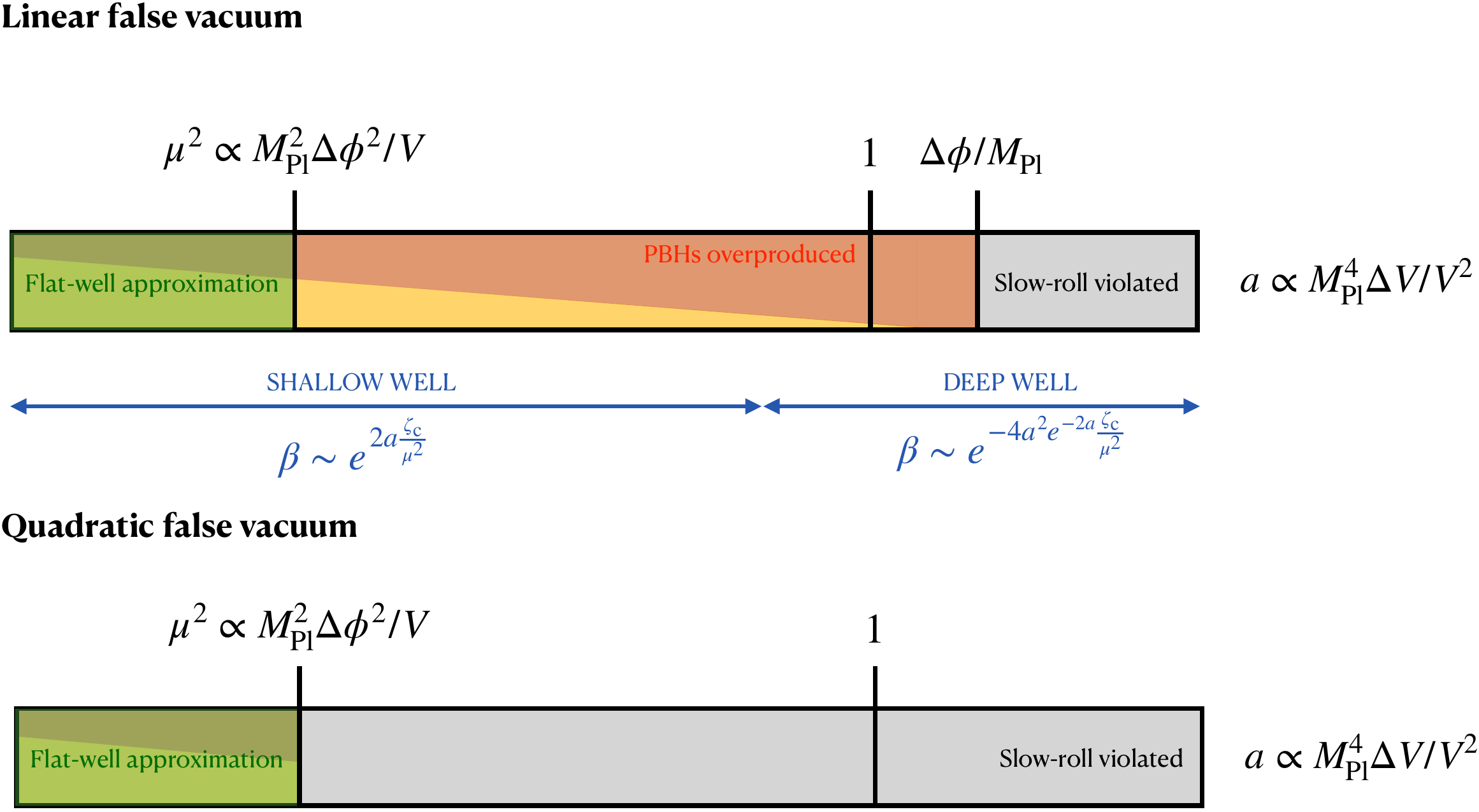}
 \caption{Summary of PBH production in false-vacuum models, which depends on two main parameters, $a$, the ratio between the height of the potential barrier and the potential squared (in Planckian units), and $\mu^2$, the ratio between the squared width of the false vacuum and the potential (still in Planckian units). In quadratic realisations of the false vacuum, the slow-roll approximation breaks down when $a>\mu^2$ (grey region), and for $a<\mu^2$ the flat-well approximation (which consists in replacing the false vacuum by a constant potential of the same width) applies (green region). In linear realisations, the slow-roll condition is less stringent and there exist a regime with large deviations from the flat-well limit, where PBHs can be overproduced (red region) or not (orange region), see main text.}
 \label{fig:summary}
\end{figure}

Despite the eminent phenomenological success of the inflationary paradigm, there are still a number of open fundamental questions that leave us far from a comprehensive understanding of the early universe. As mentioned in \Sec{sec:Introduction}, one of the main limitations is that current observations only probe a restricted range of cosmological modes, which in turn shed light on a limited interval of the inflationary phase, leaving the rest almost completely unconstrained.
The phenomenological investigation of specific features that could affect the inflationary potential at late time during inflation (hence at small scales), such as the production of primordial black holes in models featuring a local minimum, represents an interesting plan to learn more about these open issues. 

This is why in this work, we have studied cosmological fluctuations produced in potentials with a false vacuum state. Such models appear naturally when a flat inflection point is broken downwards. So by studying them one also learns about the amount of fine-tuning required in flat inflection-point models. This analysis has been done by means of the stochastic-$\delta N$ formalism, which allowed us to describe the backreaction of quantum diffusion onto the expansion dynamics of the universe. This leads to non-perturbative effects that are crucial to account for, in order to properly reconstruct the (highly non-Gaussian) tail of the distribution function of cosmic inhomogeneities. 

In practice, we have considered two toy models, depicted in \Fig{fig:potentials}, one where the potential is linear between the local minimum and the local maximum, and one where the potential is made of two quadratic profiles with opposite curvatures, arranged in such a way that the derivative of the potential vanishes at its local extremums. This has allowed us to draw some conclusions that are independent of the precise way by which the false vacuum is realised, and to highlight properties that do depend on the details of the false vacuum profile. 

We found that false-vacuum models depend on two main parameters, $a$, the ratio between the height of the potential barrier and the potential squared (in Planckian units), and $\mu^2$, the ratio between the squared width of the false vacuum and the potential energy (still in Planckian units). The mean number of \efolds elapsed while trapped in the false vacuum state depends quadratically on $\mu$ and exponentially on $a$. For the false vacuum to leave CMB scales unaffected, that number needs to be much smaller than $\sim 50$ (and potentially even smaller due to the contamination effect uncovered in \Refa{Ando:2020fjm}). Since $\mu$ is also constrained from below by the slow-roll conditions, this implies that $a$ cannot be much lager than one.

More precisely, the structure of the parameter space is summarised in \Fig{fig:summary}, and features two main regimes: the \shallow-well regime ($a\ll 1$) and the \deep-well regime ($a\gtrsim 1$). When $a\ll\mu^2$, the flat-well limit~\cite{Pattison:2017mbe}, $a=0$, provides a reliable approximation and the false vacuum can be merely described by assuming the potential to be flat between the local maximum and the local minimum. This indicates how much flat-inflection point models need to be fine tuned with respect to stochastic effects: the flat-condition breaking parameter, $a$, needs to be smaller than $\mu^2$, or in other words the potential barrier needs to be such that $\Delta V/V \ll \Delta\phi^2/\Mp^2$.

If the false vacuum is realised via a quadratic potential, this is the only regime of interest since $a>\mu^2$ leads to a violation of the slow-roll conditions (via the so-called $\eta$ parameter). In contrast, if the false vacuum is realised via a linear potential, the slow-roll conditions are less stringent and allow for the existence of a \shallow, still non-flat well regime ($\mu^2<a\ll 1$), and of a deep well regime ($a\gtrsim 1$). In the deep-well regime, PBHs are massively produced. This is due to the appearance of a new pole at the bifurcation point $a=1$,\footnote{Mathematically, we have checked that this additional pole when $a>1$ also appears in the quadratic model, although we have not reported on it since it occurs in a regime of parameters where the slow-roll approximation is violated.} which is found to depend exponentially on $a$ and which gives the decay rate of the distribution function for large cosmic inhomogeneities. In the \shallow-well regime, PBHs may be substantially produced or not, depending on the value of $\mu^2$, and the precise value of their abundance crucially depends on $a$ and on the detailed properties of the false vacuum.  

As argued above, the quadratic realisation of the false vacuum may be seen as more realistic, since it ensures that the first derivative of the potential is smoothly connected with the preceding and subsequent phases of inflation. The above considerations therefore suggest that, unless the false vacuum is so shallow that it falls into the realm of the flat-well regime, violations of slow roll are to be expected. This bears similarities with the conclusion drawn in \Refa{Ezquiaga:2019ftu} that in cubic flat-inflection point models, PBHs are overproduced unless slow roll is violated. It requires to generalise the framework developed in this work to non--slow-roll situations, which should be the topic of future investigations. 

%--------
%ACKNOWLEDGMENTS
%--------
\acknowledgments
CA thanks APC (\textit{Laboratoire Astroparticule et Cosmologie}) in Paris for the kind hospitality during the development of this work.
CA is supported in part by INFN under the program TAsP (\textit{Theoretical Astroparticle Physics}).

\appendix

\addtocontents{toc}{\protect\setcounter{tocdepth}{1}}

\section{Additional formulas}
\label{app:additional:formulas}

In this appendix, we provide additional formulas that are not given in the main text since they do not bring particular insight, but which are nonetheless useful to carry out actual computations such as those leading to the figures displayed above. 
\subsection{Characteristic function in the quadratic piecewise model}
\label{app:additional:formulas:chi}
When the integration constants $c_{\pm}$ and $d_{\pm}$ appearing in \Eq{eq:chi:quadratic:IntegrationConstantsUnfixed} are fixed using the boundary conditions~\eqref{eq:chi:BC} together with the requirement that $\chi$ and $\partial\chi/\partial\phi$ are continuous at the matching point $\phi=0$, one obtains
\bea
\label{chi_pos}
&\chi_{+}(z,x)=\frac{\hyp\left[-\frac{z^2}{a}, \frac{1}{2},a\,(x-1)^2\right]}{\hyp\left(-\frac{z^2}{a}, \frac{1}{2},a\right)}\Bigg\{\hyp\left(\frac{z^2}{a},\frac{1}{2},-a\right)-\\
& 4 \,z^4   \left[\hyp\left(\frac{1}{2}-\frac{z^2}{a}, \frac{1}{2},a\right)\hyp\left(1-\frac{z^2}{a}, \frac{3}{2},a\right)+\hyp\left(\frac{1}{2}-\frac{z^2}{a}, \frac{3}{2},a\right)\hyp\left(-\frac{z^2}{a}, \frac{1}{2},a\right)\right]\times\\
&\hyp\left(\frac{1}{2}+\frac{z^2}{a}, \frac{3}{2},-a\right) \Big/ \left[(a^2-z^2)\hyp\left(-\frac{z^2}{a}, \frac{1}{2},a\right)^2-2\, a\,(a+2z^2)\hyp\left(-\frac{z^2}{a}, \frac{1}{2},a\right)\times\right.\\
&\left.\hyp\left(-\frac{z^2}{a}, \frac{3}{2},a\right)+(a+2z^2)^2\hyp\left(-\frac{z^2}{a}, \frac{3}{2},a\right)^2\right]\Bigg\}\,,
\eea
and
\bea
\label{chi_neg}
&\chi_{-}(z,x)=e^{-a\,(1+x)^2}\,\left\{z^2\hyp\left[\frac{1}{2}-\frac{z^2}{a},\frac{1}{2},a\,(1+x)^2\right] \times\right.\\
&\left[4 \,z^2\hyp\left(1-\frac{z^2}{a},\frac{3}{2},a\right)^2-\hyp\left(-\frac{z^2}{a},\frac{1}{2},a\right) \hyp\left(1-\frac{z^2}{a},\frac{1}{2},a\right)\right.+\\
&\left. 2\, a \hyp\left(-\frac{z^2}{a},\frac{1}{2},a\right) \hyp\left(1-\frac{z^2}{a},\frac{3}{2},a\right) \right]-4 \, (1+x)\, z^4 \,\times \\
& \left[ \hyp\left(-\frac{z^2}{a},\frac{1}{2},a\right)\hyp\left(\frac{1}{2}-\frac{z^2}{a},\frac{3}{2},a\right)+ \hyp\left(\frac{1}{2}-\frac{z^2}{a},\frac{1}{2},a\right)\hyp\left(1-\frac{z^2}{a},\frac{3}{2},a\right) \right]\times\\
&\left.\hyp\left(1-\frac{z^2}{a},\frac{3}{2},a(1+x)^2\right)\right\} \Big/ \left[(a^2-z^2)\hyp\left(-\frac{z^2}{a}, \frac{1}{2},a\right)^2-2 \,a\,(a+2\,z^2)\times\right.\\
&\left. \hyp\left(-\frac{z^2}{a}, \frac{1}{2},a\right)\hyp\left(-\frac{z^2}{a}, \frac{3}{2},a\right)+(a+2\,z^2)^2\hyp\left(-\frac{z^2}{a}, \frac{3}{2},a\right)^2\right]\,.
\eea
\subsection{Mean number of \efolds in the quadratic piecewise model}
To obtain the mean number of \efolds one can apply the formula~\eqref{meanN} to the solutions for the characteristic function given by \Eqs{chi_pos} and~\eqref{chi_neg}. This leads to
\bea\label{Nplus}
\mean \N_+(x)=&\frac{\mu^2}{16\, a} \left\lbrace\pi \erf (\sqrt{a})\left[e^{2\, a}\erf (\sqrt{a}) + \erfi (\sqrt{a})\right]+\hyp^{\derhyp}\left(0,\frac{1}{2},-a\right)+ \right.\\
&\left.\hyp^{\derhyp}\left(0,\frac{1}{2},a\right)-\hyp^{\derhyp}\left[0,\frac{1}{2},a\left(x-1\right)^2\right]\right\rbrace\,,
\eea
and
\bea\label{Nminus}
\mean \N_-(x)=&\frac{\mu^2}{16 \,a} \Bigg( 4 \,a -e^{2\, a} \pi \erf (\sqrt{a})\left\lbrace\erf (\sqrt{a})-\erf\left[ \sqrt{a}\,(1+x)\right]\right\rbrace+\\
&\pi \erf \left[\sqrt{a}\,(1+x)\right]\erfi(\sqrt{a}) + a\left[\hyp^{\derhyp}\left(0,\frac{1}{2},a\right)\right]^2+\\
&a\left[-4+\hyp^{\derhyp}\left(0,\frac{3}{2},a\right)\right]\hyp^{\derhyp}\left(0,\frac{3}{2},a\right)+\\
&\hyp^{\derhyp}\left(0,\frac{1}{2},a\right)\left[1+4 \,a -2 \,a\hyp^{\derhyp}\left(0,\frac{3}{2},a\right)\right]-\\
&e^{-a\,(1+x)^2}\hyp^{\derhyp}\left[\frac{1}{2},\frac{1}{2},a\,(1+x)^2\right]-\hyp^{\derhyp}\left(1,\frac{1}{2},a\right)+\\
&2\, a\hyp^{\derhyp}\left(1,\frac{3}{2},a\right)\Bigg)\,.
\eea
In these expressions, $\erf$ and $\erfi$ denote the error function and the imaginary error function $\erfi(z)=\erf(iz)/i$ respectively, and the notation $\hyp^{\derhyp}$ stands for the derivative of the hypergeometric function with respect to its first argument. 
As a consistency check, one can verify that both expressions coincide at the matching point $\phi=0$, \ie $\mean\N$ is a continuous function of the initial condition $\phi$ as it should.

\subsection{Poles in the quadratic piecewise model}
\label{app:pole}

In this appendix, we want to solve the pole equation~\eqref{poleEq} in the shallow-well limit where $a\ll 1$. Our first step is to rewrite the pole equation in terms of Laguerre polynomials, using the identity~\cite{NIST:DLMF} 
\beq\label{Laguerre_id}
\hyp(a,b,x)=\frac{\Gamma(1-a)\,\Gamma(b)}{\Gamma(b-a)} \,\Lag(-a,b-1,x)\,,
\eeq
where $\Gamma(x)$ is the Gamma function and $\Lag(a,b,z)$ is the generalised Laguerre polynomial. This leads to
\bea
\label{eq:pole:quadratic:interm}
\frac{\pi\, z^4 \,\Gamma\left(\frac{z^2}{a}\right)^2}{ \Gamma\left(\frac{1}{2}+\frac{z^2}{a}\right)^2}\,&\left[\left(1-\frac{z^2}{a^2}\right)\Lag\left(\frac{z^2}{a},-\frac{1}{2},a\right)^2-2 \Lag\left(\frac{z^2}{a},-\frac{1}{2},a\right)\times\right.\\
&\left.\Lag\left(\frac{z^2}{a},\frac{1}{2},a\right)+\Lag\left(\frac{z^2}{a},\frac{1}{2},a\right)^2\right]=0
\eea
for the pole equation. Since the $\Gamma$ function has no zeros, the pre-factor in \Eq{eq:pole:quadratic:interm} never vanishes and the pole equation can be written as 
\bea
\label{eq:pole:quadratic:interm:2}
\left(1-\frac{z^2}{a^2}\right)\Lag\left(\frac{z^2}{a},-\frac{1}{2},a\right)^2-2 \Lag\left(\frac{z^2}{a},-\frac{1}{2},a\right)\Lag\left(\frac{z^2}{a},\frac{1}{2},a\right)+\Lag\left(\frac{z^2}{a},\frac{1}{2},a\right)^2=0\, .
\eea

In the shallow-well limit, $a\ll 1$, hence the first argument of the Laguerre polynomials is large and their last argument is small, while their product is kept fixed. Unfortunately, we are not aware of any approximation for the Laguerre polynomials in this regime, which is why we have to use a different expansion strategy. In practice, we rely on an expansion in the regime where the first argument is large,
\bea
\label{eq:Lag:exp}
\Lag(\rho,\lambda, x) = \frac{1}{\sqrt{\pi}} e^{x/2} z^{-\frac{2 \lambda+1}{4}}\rho^{\frac{2 \lambda-1}{4}} \cos\left[\left(2 \sqrt{\rho x} - \frac{\pi(2 \lambda+1)}{4}\right)\right] \left[1+\order{\frac{1}{\sqrt{\rho}}}\right],
\eea
see Eq.~(18.15.14) of \Refa{NIST:DLMF}. Note that the next term in the expansion, denoted $\order{1/\sqrt{\rho}}$, is in fact of order $1/\sqrt{\rho x}$ (see also \Refa{DEANO2013477}), so it is not further suppressed by $a$. This can therefore not be seen as an expansion in $a$, but rather as an expansion in $1/z$, which as we shall see below corresponds to an expansion in $1/n$. Inserting the above expression into \Eq{eq:pole:quadratic:interm:2}, one obtains
\bea
\label{largez}
(-3\, a+a^4-72\, z^2)\cos(4 \,z)+3 \,a\, \left[3-4\, a \,z \sin(4\, z)\right]+\mathcal{O}(1/z)=0\, .
\eea
At leading order in $1/z$, this reduces to $z^2\cos(4z)=0$. The solution $z=0$ can be safely discarded\footnote{The value $z=0$ corresponds to $t=0$, for which the characteristic function has to equal one due to the normalisation condition. This can be shown explicitly by carefully taking the limit $t\rightarrow 0$ (or equivalently $z\rightarrow 0$) in the characteristic function~\eqref{chi_pos}-\eqref{chi_neg}. This is because the numerator also vanishes at $z=0$, which is therefore not a pole.}, so $\cos(4z)=0$. This leads to $z_n= (\pi/2+ n \pi)/4$, hence
\beq
\Lambda_{n}=\frac{16}{\mu^2}z_{n}^2=\frac{\pi^2}{\mu^2}\left(n+\frac{1}{2}\right)^2+\mathcal{O}\left[(n+1/2)^0\right]\,,
\eeq
which matches the flat-well result~\cite{Pattison:2017mbe} (also obtained when letting $a=0$ in \Eq{Lambda_NW}). Since $z_n\propto n$, this justifies the above claim that the present calculation is an expansion in large $n$. In order to carry on the expansion, let us plug $z_n = (n+1/2)\pi/4 + A/(n+1/2)+ \mathcal{O}[(n+1/2)^{-2}]$ into \Eq{largez} and expand in $1/(n+1/2)$. Solving for $A$, one finds
\beq
\Lambda_{n}=\frac{16}{\mu^2}z_{n}^2=\frac{\pi^2}{\mu^2}\left[\left(n+\frac{1}{2}\right)^2+\frac{4a^2}{3\pi^2}\right]
+\mathcal{O}\left[(n+1/2)^{-1}\right]
\, .
\eeq
This expansion can be carried on again, by inserting $z_n = (n+1/2)\pi/4 + A/(n+1/2)+ B/(n+1/2)^2+ \mathcal{O}[(n+1/2)^{-3}]$ into \Eq{largez},  expanding in $1/(n+1/2)$ and solving for $B$. This leads to 
\bea
\label{Lambda_small_A}
\Lambda_n=\frac{\pi^2}{\mu^2}\left[\left(n+\frac{1}{2}\right)^2+\frac{4 a^2}{3 \pi^2 }- (-1)^n \frac{ 8 a }{\pi^3 (2 n +1) }\right]+\mathcal{O}\left[(n+1/2)^{-2}\right]\,.
\eea

\subsection{Residues in the quadratic piecewise model}
\label{app:residues}

The residues can be obtained by inserting \Eqs{chi_pos} and~\eqref{chi_neg} into \Eq{residues:alternative}. In the shallow-well limit, following the same strategy used for the pole equation, we rely on a large $z$ (and thus on a large $n$) expansion, and we can expand the corresponding expressions by making use of \Eqs{Laguerre_id} and~\eqref{eq:Lag:exp}. However, \Eq{eq:Lag:exp} applies to the case where the first argument of the Laguerre function is large positive, hence to where the first argument of the hypergeometric function is large negative. This allows us to take care of the $\chi_-$ branch of the characteristic function, since in the limit $a\ll 1$ all hypergeometric functions appearing in \Eq{chi_neg} have a large negative first argument. For the $\chi_+$ branch however, some hypergeometric functions have a large \emph{positive} first argument, such as in $\hyp(z^2/a,1/2,a)$ for instance, see \Eq{chi_pos}. This prevents us from making direct use of \Eq{eq:Lag:exp}.

This issue can be addressed by noticing that those ``problematic'' terms do not depend on $x$. More precisely, \Eq{chi_pos} can be rewritten as 
\bea
\chi_+(z,x)=\frac{h(z,x)}{h(z,0)}j(z)\, ,
\eea
where
\bea
h(z,x)=\hyp\left[-\frac{z^2}{a},\frac{1}{2}, a(x-1)^2\right]
\eea
and $j(z)$ is the quantity in braces in \Eq{chi_pos}. The ``problematic'' terms, \ie those containing hypergeometric functions with a first argument that is positive in the regime $a\ll 1$, are contained in $j(z)$. The trick is to use the continuity of the characteristic function at the matching point $x=0$, which implies that
\bea
j(z)=\chi_-(z,0)\, .
\eea
This leads to an expression for $\chi_+(z,x)$,
\bea
\chi_+(z,x)=\frac{h(z,x)}{h(z,0)}\chi_-(z,0)\, ,
\eea
which does not involve any ``problematic'' term, since $\chi_-$ does not contain any such term as mentioned above. This allows us to expand both branches of the characteristic function using \Eq{eq:Lag:exp}.

In the $\chi_+$ branch, this leads to
\bea
a_n^{(+)}(x)\simeq &\frac{e^{-a\,+\frac{1}{2}\,a\,(x-1)^2}}{6 \mu^2 \left[(3+a^2)\cos{4 z_n}-6 z_n \sin{4 z_n}\right]}\Big\{-36 \,a \cos{\left[2 \,(x-3)\,z_n\right]}\, +\\
&\left[-6 \,a^3 (x-1)^4+a^4(x-1)^6-9\, a^2(x-3)(x+1)-288\, z_n^2 + 72 \,a \cos{4 z_n}\right]\times\\
&\cos{\left[2\,(x-1)\,z_n\right]}-36 \,a \cos{\left[2 \,(x+1)\,z_n\right]}+24 \,a \left[-3+a(x-1)^2\right]\times\\
&(x-1)\,z_n \sin{\left[2\,(1-x)\,z_n\right]}\,\Big\}\, .
\eea
By replacing $z_n$ by their expression found in \App{app:pole}, namely $z_n = (n+1/2)\pi/4 + A/(n+1/2)+ B/(n+1/2)^2+ \mathcal{O}[(n+1/2)^{-3}]$ and further expanding in $1/(n+1/2)$, one finds 
\beq\label{res_pos_quad:app}
\begin{split}
a_n^{{+}}(x)=&\frac{(-1)^n e^{\frac{1}{2}a[x(x-2)-1]}}{\mu^2}\left(2 \pi \left(n+\frac{1}{2}\right)\cos{\left[\frac{\pi}{2}\left(n+\frac{1}{2}\right)(x-1)\right]}+\right.\\
&f_1(a,x) \sin{\left[\frac{\pi}{2}\left(n+\frac{1}{2}\right)(x-1)\right]}+\\
&\frac{1}{\pi}\left(n+\frac{1}{2}\right)^{-1}\left\lbrace4\, a\cos{\left[\frac{\pi}{2}\left(n+\frac{1}{2}\right)(x-3)\right]} +4\, a \cos{\left[\frac{\pi}{2}\left(n+\frac{1}{2}\right)(x+1)\right]}+\right.\\
&\left.\left.a^2 f_2(a,x) \cos{\left[\frac{\pi}{2}\left(n+\frac{1}{2}\right)(x-1)\right]}\right\rbrace+\mathcal{O}(n+1/2)^{-2}\right)\,,
\end{split}
\eeq
where
\bea
f_1(a,x)=&\frac{2}{3}\, a (x-1) [a(x-2)x-3]\,,\\
f_2(a,x)=&\frac{1}{9}\left(-39 + 9(x-2)x + 6 ax (x-2)(x-1)^2-\right.\\
&\left.a^2\{7+(x-2)x[x(x-1)^2(x-2)-1]\}\right)\,.
\eea

A similar calculation can be performed in the $\chi_-$ branch. However, in the main text we are mostly interested in the PDF of the number of \efolds when starting from the bottom of the potential, $x=1$, which makes this expression of limited interest. This is why we do not display here, although it can be readily derived using the above formulas.

\bibliographystyle{JHEP}
\bibliography{biblio}
\end{document}